\documentclass[12pt]{article}
\usepackage{amsmath}
\usepackage{amssymb}
\usepackage{graphicx}
\usepackage{natbib}
\usepackage{url} 
\usepackage{booktabs}
\usepackage{multirow}
\usepackage{setspace}
\usepackage{comment}
\usepackage{float}

\newcommand{\blind}{0}

\newcommand{\bbeta}{\mbox{\boldmath $\beta$}}

\newcommand{\hatbbeta}{\mbox{$\boldsymbol{\skew{2}\hat\beta}$}}

\newcommand{\indep}{\perp\!\!\!\perp}
\DeclareSymbolFont{matha}{OML}{txmi}{m}{it}
\DeclareMathSymbol{\varv}{\mathord}{matha}{118}

\DeclareMathOperator*{\argmin}{arg\,min}

\newtheorem{prop}{Proposition}
\newtheorem{theorem}{Theorem}
\newtheorem{lemma}{Lemma}

\begin{document}

\def\spacingset#1{\renewcommand{\baselinestretch}%
{#1}\small\normalsize} \spacingset{1}


\if0\blind
{
  \title{\bf Targeted Parameter Estimation for Robust Empirical Bayes Ranking}
  \author{Nicholas C. Henderson
    \hspace{.2cm}\\
    Department of Biostatistics, University of Michigan\\
    and \\
    Nicholas Hartman \\
    Department of Biostatistics, University of Michigan
    }
    \date{}
  \maketitle
} \fi

\if1\blind
{
  \bigskip
  \bigskip
  \bigskip
  \begin{center}
    {\LARGE\bf Targeted Parameter Estimation for Robust Empirical Bayes Ranking}
\end{center}
  \medskip
} \fi

\bigskip
\begin{abstract}
Ordering the expected outcomes across a collection of clusters after performing a covariate adjustment commonly arises in many applied settings, such as healthcare provider evaluation.
Regression parameters in such covariate adjustment models are frequently
estimated by maximum likelihood or through other criteria that do not
directly evaluate the quality of the rankings resulting from using 
a particular set of parameter estimates. 
In this article, we propose both a novel empirical Bayes ranking procedure 
and an associated estimation approach for finding the regression parameters
of the covariate adjustment model. By building our ranking approach around estimating approximate percentiles of the covariate-adjusted cluster-level means, we are able to develop manageable expressions for the expected ranking squared-error loss associated with any choice of the covariate-adjustment model parameters, and we harness this to generate a novel unbiased estimator for this expected loss. Minimization of this unbiased estimator directly leads to
a novel ranking procedure that is often more robust than conventional empirical Bayes ranking methods. 
Through a series of simulation studies, we show that our approach consistently delivers improved ranking squared-error performance relative to competing methods, such as posterior expected ranks and ranking the components of the best linear unbiased predictor. Estimating rankings using our method is illustrated with 
an example from a longitudinal study evaluating test scores across a large group of schools.
\end{abstract}

\noindent%
\vfill

\newpage

\spacingset{1.65}

\section{Introduction}
\label{sec:intro}

Comparing average outcomes across different clusters is an objective that arises in a wide range of applied contexts. Examples include evaluation of hospital performance \citep{ spiegelhalter2012, normand2016, wu2021}, comparison of value-added performance of schools on student outcomes \citep{leckie2024}, ordering anti-cancer drugs according to their efficacy in cell lines \citep{gerdes2021}, and ranking genetic locations in two-group comparisons made in high-throughput biological experiments \citep{ferguson2020}. Variation in expected outcomes across such clusters, which we refer to as ``units'', is often driven, in part, by differences in covariates summarized at the unit level. For example, differences in the case mix across hospitals are frequently a major driver of the variation in observed mortality rates. However, observed variation in unit-level mean outcomes is often not sufficiently explained by the variation in unit-level covariates, and when this occurs, there is often interest in comparing unit-level means after an adjustment for available covariates has been performed. Such comparisons can facilitate the identification of units with especially low or high outcomes relative to what one would expect, given the observed attributes of that unit \citep{jones2011, mcculloch2024}. Ranking is a useful framework for assessing the relative positioning of different units and for detecting units with extreme mean values in a way that is robust to the measurement scale used.

The problem of sorting a collection of units according to the mean of outcomes within those units has been a well-studied problem in statistics \citep{gibbons1979} and has led to the development of a broad range of statistical methods \citep{gu2023}. Among such methods, approaches that could be labeled as empirical Bayes, shrinkage estimation, or best prediction from linear mixed models, where unit-level means are viewed as draws from a common distribution, have played a particularly prominent role \citep{shen1998}. A chief advantage of shrinkage estimation approaches is the way that they seamlessly handle settings with high estimation variance and substantial variability in estimation variance across units \citep{henderson2016}. This is because estimates for units with high estimation variance are shrunken towards a common mean or regression surface, which results in highly variable units being less likely to be highly ranked when compared to ranking the direct estimates of the unit-level means. In addition,
when compared to ranking by p-values associated with unit-level hypothesis tests, empirical Bayes procedures are less prone to highly ranking units with very low estimation variance and modest effect size estimates \citep{gelman1999}.

Empirical Bayes ranking procedures are typically obtained by employing two-level hierarchical models, where the form of the ranking rule is found by either ranking more familiar Bayes rules, such as posterior means, or by deriving a Bayes rule that is optimal with respect to a ranking-specialized loss function \citep{laird1989, noma2013}. The final estimated rankings are then obtained by plugging in estimates of any remaining unknown parameters into the ranking rule derived from the assumed two-level hierarchical model.
While plugging in likelihood-based parameter estimates often leads to rankings with good operating characteristics, 
the likelihood function does not directly incorporate a criterion that measures the performance
of the empirical Bayes rankings resulting from using these likelihood-based estimates \citep{xie2012}. 
Integrating a ranking-based objective function
into parameter estimation has the potential to yield superior performance, particularly in the presence of model misspecification.

When the main interest is in estimating the unit-level means, the best linear unbiased predictor (BLUP) arising from a model-based linear mixed model is a well-studied and effective approach \citep{robinson1991}. However, as noted by \cite{jiang2011}, when the linear model for the fixed effects is misspecified, substantial improvements in the squared prediction error for the unit-level parameters can often be made by utilizing an objective function that directly targets this mean-squared prediction error instead of using likelihood-based estimates of the fixed effects regression parameters. This occurs because the BLUP estimates are obtained by shrinking unit-specific direct estimates towards a regression-based prediction for each unit, and units with low estimation variance play an outsized role in determining the form of this regression prediction. Under model misspecification, improvements in mean-squared error can be achieved by minimizing an objective function where units with low estimation variance receive less weight in the estimation of the regression function. In this paper, we adopt a similar framework as \cite{jiang2011} and extend it to the ranking setting, where we develop a novel approach to parameter estimation when the objective is to rank a collection of unit-level attributes. Specifically, we propose to estimate the fixed effects regression coefficients by minimizing an objective function that targets the squared error between an estimated ranking and the ranking of the underlying covariate-adjusted unit-level means. This approach can generate notable improvements in ranking-based squared error, particularly in cases with substantial heteroskedasticity in estimation variance and some degree of model misspecification.

To construct an objective function that targets both ranking squared-error loss and computational tractability, we focus on squared-error loss for parameters that we refer to as population percentiles, rather than squared-error loss for the percentiles of the mean parameters (i.e., a loss that is a normalized version of ranking squared-error loss). Using these population percentiles as the objects of interest, we show that the best decision rule for estimating population percentiles is to use the posterior expected population percentiles, which have a simple, direct form. The direct form of the posterior expected population percentiles enables us to obtain a tractable form for their associated risk under squared-error loss. Building on this, we then develop a novel, unbiased estimator of this risk, which directly leads to the construction of an estimator of the fixed effects regression coefficients. These regression coefficient estimates can then be plugged into the posterior expected population percentiles to generate the final collection of percentile estimates.

This paper is organized as follows. Section 2 describes the data structure of interest, introduces key model notation, and defines the percentile parameters of interest and their population percentile analogs. This section then introduces a novel percentiling procedure that is optimal for estimating the population percentiles when the fixed effects regression coefficients are known. Section 3 describes an unbiased estimator for the expected loss associated with the percentiling procedure developed in Section 2, and this section then details how minimization of this unbiased risk estimator can be used to generate percentile estimates from the observed data. Section 3 also introduces a novel majorize-minimize (MM) algorithm for performing minimization of our proposed unbiased risk estimator. Section 4 presents the results of three simulation studies that evaluate and compare our approach with several competing methods. Section 5 illustrates the use of our percentiling method with an application to a longitudinal study evaluating educational effectiveness, and we conclude with a brief discussion in Section 6.
\vspace{-0.5cm}

\section{Statistical Model and Evaluating Percentiling Procedures}
\label{sec:model}
\subsection{Statistical Model}
We assume that we have observed $K$ summary statistics $Y_{1}, \ldots, Y_{K}$ that correspond to 
$K$ separate clusters or ``units''.  Each summary statistic $Y_{k}$ is an estimate
of an associated parameter $\theta_{k}$, which represents
an attribute of interest from unit $k$. For example, $Y_{k}$ could represent a sample
mean of individual observations within unit $k$, a Horvitz-Thompson estimate, or an estimated $t$-year survival probability.
For all units, we also have standard errors $\sigma_{1}, \ldots, \sigma_{K}$, where $\sigma_{k}$ estimates the standard deviation of $Y_{k}$. In addition to $(Y_{k}, \sigma_{k})$,
all units have associated $p \times 1$ covariate vectors
$\mathbf{x}_{1}, \ldots, \mathbf{x}_{K}$ whose purpose is to explain
part of the variation in the parameters of interest $\theta_{k}$. 

We assume the distribution of $Y_{k}$ is determined by the following model
\begin{eqnarray}
Y_{k} &=& \theta_{k} + e_{k} \nonumber \\
\theta_{k} &=& \mu_{k} + \varv_{k}, \label{eq:true_model}
\end{eqnarray}
where $\varv_{k} \sim N(0, \tau^{2})$, $e_{k} \sim N(0, \sigma_{k}^{2})$, $\varv_{k}$ and $e_{k}$ are independent. In this model, the sampling error in the standard errors is not accounted for, and the values of $\sigma_{k}^{2}$ are assumed to be fixed and ``known''. Moreover, $Y_{1}, \ldots, Y_{K}$ are assumed to be mutually independent.
In (\ref{eq:true_model}), $\mu_{k}$ represents the fixed mean of unit $k$, where, unconditional on $\varv_{k}$, $E( Y_{k} ) = \mu_{k}$. The fixed means $\mu_{k}$ in models such as (\ref{eq:true_model}) are often assumed to be a function of the unit-level covariates $\mathbf{x}_{k}$, but we do not assume here that the $\mu_{k}$ have any specific relationship with the $\mathbf{x}_{k}$.

Our main goal is to estimate the ranking of the unobserved random effects $\varv_{k}$, where the ranking is obtained by sorting the units in ascending order.  
Specifically, the rank of each random effect $\varv_{k}$ is defined as
\begin{equation}
\textrm{rank}(\varv_{k}; \mathbf{v}_{-k}) = \sum_{j=1}^{K} I(\varv_{k} \geq \varv_{j}), \nonumber
\end{equation}
where $\mathbf{v} = (\varv_{1}, \ldots, \varv_{K})^{T}$ is the $K \times 1$ vector of random effects and $\mathbf{v}_{-k}$ is the vector containing all random effects except for $\varv_{k}$.
In the absence of ties, the largest value of $\varv_{k}$ is assigned a rank of $K$, while the smallest $\varv_{k}$ is assigned a rank of $1$. It is often more convenient to work with the ``normalized percentiles'' of $\varv_{k}$ rather than the rankings \citep{Lockwood2002}. The normalized percentile $\textrm{perc}(\varv_{k};\mathbf{v}_{-k})$ of $\varv_{k}$ is defined as
\begin{equation}
\textrm{perc}(\varv_{k}; \mathbf{v}_{-k}) = \frac{\textrm{rank}(\varv_{k}; \mathbf{v})}{K+1} = \frac{1}{K+1}\sum_{j=1}^{K} I(\varv_{k} \geq \varv_{j}). \nonumber
\end{equation}

A common approach (e.g., \cite{fay1979}) to modeling the $\mu_{k}$ is to use a linear combination of the covariates, i.e., $\mu_{k} = \mathbf{x}_{k}^{T}\bbeta$. We will refer to the model that makes the working assumption that $\mu_{k} = \mathbf{x}_{k}^{T}\bbeta$ as the ``working model'' to distinguish it from the true data generating model defined in (\ref{eq:true_model}). Specifically, we define the working model as
\begin{equation}
\textrm{working model: } \begin{cases} 
Y_{k} = \mathbf{x}_{k}^{T}\bbeta + \varv_{k} + e_{k},\qquad 
k = 1, \ldots, K,  \\
\varv_{k} \sim N(0, \tau^{2}), \quad e_{k} \sim N(0, \sigma_{k}^{2}), \quad \varv_{k} \indep e_{k},
\end{cases}
\label{eq:working_model}
\end{equation}
where $\varv_{k} \indep e_{k}$ denotes that $\varv_{k}$ and $e_{k}$ are independent.

For fixed $\bbeta$ and $\tau$, the best predictor (BP) \citep{searle2009} of $\varv_{k}$ under the working 
model (\ref{eq:working_model}) is the expectation of $\varv_{k}$ conditional on $Y_{k}$, which is given by
\begin{equation}
\textrm{BP}_{k} = E_{W}\{ \varv_{k} \mid Y_{k} \}
= B_{k,\tau}(Y_{k} - \mathbf{x}_{k}^{T}\bbeta),
\end{equation}
where $E_{W}\{ \}$ denotes taking expectation assuming working model (\ref{eq:working_model}) holds and where the shrinkage factor $B_{k,\tau}$ is defined as $B_{k,\tau} = \tau^{2}/(\tau^{2} + \sigma_{k}^{2})$.
It is well known that the best predictor minimizes the mean-squared error when aiming to estimate the random 
effect $\varv_{k}$ \citep{mcculloch2004}.
However, under many standard loss functions used for quantifying the quality of rankings (e.g., \cite{Lin2006, jewett2019}), ranking by the values of $\textrm{BP}_{k}$ is often a sub-optimal way to rank the random effects. Indeed, under squared-error loss for ranks or percentiles, one can directly show that the optimal estimates of the percentiles of the $\varv_{k}$ are the posterior expected percentiles, where the posterior is obtained by considering the assumed Gaussian distribution of $\varv_{k}$ to be a prior distribution.  
In our setting, where $(Y_{k}, \varv_{k})$ has the joint distribution defined by (\ref{eq:true_model}), the posterior expected percentile (PEP) of $\varv_{k}$ is given by
\begin{eqnarray}
\textrm{PEP}_{k} &=& E\{ \textrm{perc}(\varv_{k}; \mathbf{v}_{-k}) \mid \mathbf{Y} \} \nonumber \\
&=& \frac{1}{K+1} + \frac{1}{K+1}\sum_{j \neq k} \Phi\Big( \frac{B_{k,\tau}(Y_{k} - \mu_{k}) - B_{j,\tau}(Y_{j} - \mu_{j} )}{ B_{k,\tau}\sigma_{k}^{2} + B_{j,\tau}\sigma_{j}^{2} }  \Big),
\label{eq:pep_full}
\end{eqnarray}
where $\mathbf{Y}$ is the $K \times 1$ vector $\mathbf{Y} = (Y_{1}, \ldots, Y_{K})^{T}$ and
$\Phi(\cdot)$ denotes the cumulative distribution function of a standard normal random variable.

\subsection{Population Percentiles}
The percentile of the $k^{th}$ unit $\textrm{perc}(\varv_{k}; \mathbf{v}_{-k})$
depends on the entire vector of random effects $\mathbf{v}$, and consequently, the posterior expected percentiles depend on the entire vector $\mathbf{Y}$. This dependence on $\mathbf{Y}$ and resulting form of $\textrm{PEP}_{k}$ in 
(\ref{eq:pep_full}) has two main drawbacks. First, when compared to $\textrm{BP}_{k}$, the roles that the residual $Y_{k} - \mu_{k}$,
the standard error $\sigma_{k}$, and the random effect standard deviation $\tau$ play in determining the form of $\textrm{PEP}_{k}$ are not transparent and can be difficult to ascertain from examining Equation (\ref{eq:pep_full}). Secondly, the form of $\textrm{PEP}_{k}$ in (\ref{eq:pep_full}) will be challenging to utilize when developing an objective function that targets estimation of $\bbeta$ in order to improve the ranking of the $\varv_{k}$. 

A close approximation of $\textrm{perc}(\varv_{k}; \mathbf{v}_{-k})$ that removes the dependence of $\textrm{perc}(\varv_{k}; \mathbf{v}_{-k})$ on the entire vector $\mathbf{v}$ and leads to a more direct form for an approximate posterior expected percentile is
\begin{eqnarray}
\rho_{k} = P(\varv_{k} \geq \varv_{j} \mid \varv_{k}) =  F_{v}(\varv_{k}) = \Phi(\varv_{k}/ \tau),
\end{eqnarray}
where $F_{v}(t) = P(\varv_{j} \leq t) = \Phi(t/\tau)$ is the cumulative distribution function of $\varv_{j}$. We define $\rho_{k}$ to be the population percentile
of unit $k$ because $\textrm{perc}(\varv_{k}; \mathbf{v}_{-k})$ will converge in probability to $\rho_{k}$ as $K \longrightarrow \infty$, and thus, $\rho_{k}$ can be thought of as the percentile one would assign to unit $k$ in an ``infinite unit'' setting.

One can directly show (see Appendix A) that, under working model (\ref{eq:working_model}), the posterior expected population percentile (PEPP) $R_{k}(\bbeta, \tau, Y_{k})$ of $\varv_{k}$ is given by
\begin{equation}
R_{k}(\bbeta, \tau, Y_{k}) = E_{W}\big\{ \rho_{k} \big|  \bbeta, \tau^{2}, Y_{k} \big\} = \Phi\Big( V_{k,\tau}(Y_{k} - \mathbf{x}_{k}^{T}\bbeta) \Big),
\label{eq:pep_definition}
\end{equation}
where the term $V_{k,\tau}$ in (\ref{eq:pep_definition}) is defined as
\begin{equation}
V_{k,\tau} = \sqrt{B_{k,\tau}/(2\sigma_{k}^{2} + \tau^{2})}.
\label{eq:ranking_shrink}
\end{equation}
Note that, unlike the shrinkage terms $B_{k,\tau}$, the $V_{k,\tau}$ are not constrained to lie between $0$ and $1$; instead, the $V_{k,\tau}$ are guaranteed to satisfy $0 \leq V_{k,\tau} \leq 1/\tau$.

One should also note that the $R_{k}(\bbeta, \tau, Y_{k})$ are not themselves ``proper percentiles'' in the sense that they do not fully populate the values $\tfrac{1}{K+1}, \ldots, \tfrac{K}{K+1}$, and they should be thought of as only estimates of the underlying percentiles $\textrm{perc}(\varv_{k}, \mathbf{v}_{-k})$ of the random effects, not as percentiles themselves.
Moreover, percentiles obtained by ranking $R_{k}(\bbeta, \tau, Y_{k})$ will,
in general, not be the same as the percentiles one would obtain from ranking the values of $\textrm{BP}_{k}$. This can be seen by noting the following relationship between $R_{k}(\bbeta, \tau, Y_{k})$ and $\textrm{BP}_{k}$
\begin{equation}
R_{k}(\bbeta, \tau, Y_{k}) = \Phi\Big( \textrm{BP}_{k}\sqrt{ \frac{\tau^{2}(\sigma_{k}^{2} + \tau^{2})}{ 2\sigma_{k}^{2} + \tau^{2}} } \Big). \nonumber
\end{equation}
Through the multiplication of $\textrm{BP}_{k}$ by $\{\tau^{2}(\sigma_{k}^{2} + \tau^{2})/(2\sigma_{k}^{2} + \tau^{2})\}$ in the above, the variability in the standard errors $\sigma_{k}$ can, in many settings, generate an ordering of $R_{k}(\bbeta, \tau, Y_{k})$ that is different from the $\textrm{BP}_{k}$ ordering.

\noindent
\textit{Posterior expected population percentiles with individual-level data.}
If one has individual-level observations $Y_{kj}$ within unit $k$ instead of only unit-level summary statistics $Y_{k}$,
a working model assumption analogous to (\ref{eq:working_model}) is $Y_{kj} = \mathbf{x}_{kj}^{T}\bbeta + \varv_{k} + e_{kj}$, where $\varv_{k} \sim N(0, \tau^{2})$ and $e_{kj} \sim N(0, \sigma^{2})$. In this case, the posterior expectation of the population percentile $\rho_{k} = \Phi(\varv_{k}/\tau)$ will be 
$E\{\rho_{k} \mid Y_{k1}, \ldots, Y_{kn_{k}} \} = \Phi\big( \tilde{V}_{k,\tau}(\bar{Y}_{k} - \bar{\mathbf{x}}_{k.}^{T}\bbeta) \big)$,
where $\bar{Y}_{k.} = \tfrac{1}{n_{k}}\sum_{j=1}^{n_{k}} Y_{kj}$, 
$\bar{\mathbf{x}}_{k.} = \tfrac{1}{n_{k}}\sum_{j=1}^{n_{k}} \mathbf{x}_{kj}$,
and $\tilde{V}_{k,\tau} = \tau \{ (\tau^{2} + \sigma^{2}/n_{k})(2\sigma^{2}/n_{k} + \tau^{2}) \}^{-1/2}$.
In this setting, the PEPP has the same form as (\ref{eq:pep_definition}) if one replaces $Y_{k}$ with $\bar{Y}_{k.}$, replaces $\mathbf{x}_{k}$ with $\bar{\mathbf{x}}_{k.}$, and replaces $\sigma_{k}^{2}$ with $\sigma^{2}/n_{k}$.

\subsection{Loss functions evaluating percentiling methods}
The posterior expected percentiles defined in (\ref{eq:pep_full}) are the optimal percentiles to report under squared-error loss for percentiles.
For a collection of data-dependent percentiles $A_{1}(\bbeta, \tau, Y_{1})$, ..., $A_{K}(\bbeta, \tau, Y_{K})$ that can depend on $(\bbeta, \tau)$, the percentile squared-error loss (PSEL) is defined as
\begin{equation}
\textrm{PSEL}_{\tau}(\bbeta; A_{1}, \ldots, A_{K}) = 
 \frac{1}{K}\sum_{k=1}^{K} \Big( \textrm{perc}(\varv_{k}; \mathbf{v}_{-k}) - A_{k}(\bbeta, \tau, Y_{k}) \Big)^{2}. 
 \label{eq:psel_def}
\end{equation}
One can note that the PSEL is simply a normalized version of ranking squared-error loss.

As a close approximation to the PSEL in (\ref{eq:psel_def}), we also introduce the population percentile squared-error (PPSEL), which replaces the parameter percentiles $\textrm{perc}(\varv_{k}; \mathbf{v}_{-k})$ with their population versions $\rho_{k}$. For percentiles $A_{k}(\bbeta, \tau, Y_{k})$, the PPSEL is defined as
\begin{equation}
\textrm{PPSEL}_{\tau}(\bbeta; A_{1}, \ldots, A_{K}) = 
 \frac{1}{K}\sum_{k=1}^{K} \Big( \rho_{k} - A_{k}(\bbeta, \tau, Y_{k}) \Big)^{2}. \nonumber 
\end{equation}
The expectation of $\textrm{PPSEL}_{\tau}(\bbeta; A_{1}, \ldots, A_{K})$, under working model (\ref{eq:working_model}),
is minimized by setting $A_{k}(\bbeta, \tau, Y_{k})$ to the PEPP $R_{k}(\bbeta, \tau, Y_{k})$. While the $R_{k}(\bbeta, \tau, Y_{k})$ are optimal for the expected PPSEL when $\bbeta$ is fixed and when working model (\ref{eq:working_model}) holds, the fixed-effects regression coefficients $\bbeta$ must be estimated in practice.
The performance of $R_{k}(\bbeta, \tau, Y_{k})$ can differ meaningfully across various estimates of $\bbeta$ that one might plug into $R_{k}(\bbeta, \tau, Y_{k})$, and such differences can be impacted by the degree to which the assumptions of
working model (\ref{eq:working_model}) fail to hold.

\vspace{-0.5cm}

\section{Ranking-targeted Estimation of Fixed Effects and ROPPER}
\label{s:inf}

\subsection{Unbiased Estimation of PPSEL}
One can think of the posterior expected population percentiles $R_{k}(\bbeta, \tau, Y_{k})$ in (\ref{eq:pep_definition})
as defining a family of percentile estimates indexed by $\bbeta$, and our aim is to find the values of $\bbeta$ that optimize performance within this class of percentile estimators.
While one possible choice is to plug in the maximum likelihood estimate of $\bbeta$ into $R_{k}(\bbeta, \tau, Y_{k})$, a potentially more effective choice is to consider the expected PPSEL associated with using $R_{k}(\bbeta, \tau, Y_{k})$. This expected loss can be expressed as
\begin{eqnarray}
Q_{\tau}(\bbeta) &=&  E\Big\{ \textrm{PPSEL}_{\tau}(\bbeta; R_{1}, \ldots, R_{K}) \Big\} \nonumber \\
&=& \frac{1}{12} - \frac{2}{K}\sum_{k=1}^{K}E\Big\{ (\rho_{k} - 1/2)R_{k}(\bbeta, \tau, Y_{k}) \Big\}
+ \frac{1}{K}\sum_{k=1}^{K} E\Big\{ \Big( R_{k}(\bbeta, \tau, Y_{k}) - \frac{1}{2}\Big)^{2} \Big\}.
\label{eq:Qfn}
\end{eqnarray}
To find a value of $\bbeta$ that targets the expected loss $Q_{\tau}(\bbeta)$ directly, we propose to minimize an unbiased estimate of $Q_{\tau}(\bbeta)$. Estimation by minimizing an unbiased risk estimator has been widely used in hyperparameter tuning and as an effective technique in empirical Bayes shrinkage methods (e.g., \cite{xie2012, rosenman2023})

The unbiased estimate of the risk $Q_{\tau}(\bbeta)$ that we propose to use for estimation of $\bbeta$ is based on the following estimator
\begin{eqnarray}
\hat{Q}_{\tau}(\bbeta) 
&=& \frac{1}{12} - \frac{1}{K}\sqrt{\frac{2}{\pi}}\sum_{k=1}^{K} \sum_{h=0}^{\infty} \frac{(-1)^{h}}{2^{h}h!(2h+1)}\sum_{j=0}^{h} \frac{(2h + 1)!(\tau^{2})^{h - j + 1/2}}{2^{j}(2h + 1 - 2j)!j!}\Big[ \frac{\partial^{2h - 2j + 1} R_{k}(\bbeta, \tau, Y_{k})}{\partial Y_{k} } \Big] \nonumber \\
&+&  \frac{1}{K}\sum_{k=1}^{K} \Bigg(\Phi\Big( V_{k,\tau}(Y_{k} - \mathbf{x}_{k}^{T}\bbeta) \Big) - \frac{1}{2} \Bigg)^{2}.
\label{eq:Qhat}
\end{eqnarray}
An important thing to note is that, for a fixed $\bbeta$, $\hat{Q}_{\tau}(\bbeta)$ does not depend on
the unknown means $\mu_{k}$ in any way. Moreover, while $\hat{Q}_{\tau}(\bbeta)$ measures the risk associated with the percentiles obtained from working model (\ref{eq:working_model}), $\hat{Q}_{\tau}(\bbeta)$ is an unbiased estimator of $Q_{\tau}(\bbeta)$, where the expectation is taken under the true model (\ref{eq:true_model}).
This result is stated in Theorem \ref{thm:unbiased_estimate} below.
\vspace{-0.3cm}
\begin{theorem}
Under model (\ref{eq:true_model}), $\hat{Q}_{\tau}(\bbeta)$ is an unbiased estimator of $Q_{\tau}(\bbeta)$ in the sense that,
for any $\tau > 0$ and $\bbeta \in \mathbb{R}^{p}$,
\begin{equation}
E\{ \hat{Q}_{\tau}(\bbeta) \} = Q_{\tau}(\bbeta).  \nonumber
\end{equation}
\label{thm:unbiased_estimate}
\end{theorem}

\vspace{-0.5cm}

As can be seen in (\ref{eq:Qhat}), one of the terms in $\hat{Q}_{\tau}(\bbeta)$ involves an infinite summation, and in practice, one will need to use a finite sum as an approximation of this infinite sum.
We define the order-H approximation, for $H \geq 1$, of the unbiased risk estimate $\hat{Q}_{\tau}(\bbeta)$ as
\begin{eqnarray}
\hat{Q}_{\tau}^{(H)}(\bbeta) 
&=& \frac{1}{12} - \frac{1}{K}\sqrt{\frac{2}{\pi}}\sum_{k=1}^{K} \sum_{h=0}^{H-1} \frac{(-1)^{h}}{2^{h}h!(2h+1)}\sum_{j=0}^{h} \frac{(2h + 1)!(\tau^{2})^{h - j + 1/2}}{2^{j}(2h + 1 - 2j)!j!}\Big[ \frac{\partial^{2h - 2j + 1} R_{k}(\bbeta, \tau, Y_{k})}{\partial Y_{k} } \Big] \nonumber \\
&+&  \frac{1}{K}\sum_{k=1}^{K} \Bigg(\Phi\Big( V_{k,\tau}(Y_{k} - \mathbf{x}_{k}^{T}\bbeta) \Big) - \frac{1}{2} \Bigg)^{2}. \nonumber 
\end{eqnarray}

We have experimented with many choices of $H$, and in our experience, the first-order approximation (where $H=1$) works very well in practice. There are typically minor differences in performance between the first-order approximation and approximations using higher orders (see Appendix E).  
The order-1 approximation $\hat{Q}^{(1)}(\bbeta)$ can be simplified to
\begin{equation}
\hat{Q}_{\tau}^{(1)}(\bbeta) 
= \frac{1}{12} - \frac{\tau}{K}\sqrt{\frac{2}{\pi}}\sum_{k=1}^{K} V_{k,\tau}\phi\Big( V_{k,\tau}(Y_{k} - \mathbf{x}_{k}^{T}\bbeta) \Big) +  \frac{1}{K}\sum_{k=1}^{K} \Bigg(\Phi\Big( V_{k,\tau}(Y_{k} - \mathbf{x}_{k}^{T}\bbeta) \Big) - \frac{1}{2} \Bigg)^{2}. \nonumber 
\end{equation}

We refer to the fixed effects estimates $\hatbbeta_{r}$ obtained by minimizing $\hat{Q}^{(H)}(\bbeta)$ as the ranking-focused unbiased risk estimate (RFURE) of the regression coefficients $\bbeta$. The percentiles obtained by plugging $\hatbbeta_{r}$ into the posterior expected population percentile functions defined by (\ref{eq:pep_definition}) are given by $R_{1}(\hatbbeta_{r}, \tau, Y_{k}), \ldots, R_{K}(\hatbbeta_{r}, \tau, Y_{K})$. We refer to this collection of percentile estimates as the Ranking-Optimized Population Posterior Expected peRcentiles (ROPPER).

\subsection{Estimating the random effects variance $\tau^{2}$}

The unbiased risk estimator $\hat{Q}_{\tau}(\bbeta)$ estimates the expected PPSEL when $\tau$ is assumed to be fixed,
and one should only minimize $\hat{Q}_{\tau}(\bbeta)$ with respect to $\bbeta$ and not perform joint minimization with respect to $(\bbeta, \tau)$. This is because population percentiles $\rho_{k}$ are functions of $\tau$, and an estimate of $\bbeta$ that minimizes PPSEL should only be interpreted as an optimal estimate under the assumption that the random effects have variance $\tau^{2}$. That is, one should not compare the PPSEL across different values of $\tau$; hence, using $\hat{Q}_{\tau}(\bbeta)$ to guide estimation of $\tau$ is not appropriate.
For this reason, our approach is to first compute an estimate $\hat{\tau}$ of $\tau$ using a method that aims to accurately estimate the random effects standard deviation, and using this estimate of $\tau$, we find $\hatbbeta_{r}$ by minimizing $\hat{Q}_{\tau}(\bbeta)$ with $\tau$ set equal to $\hat{\tau}$.

We have evaluated the performance of ROPPER using two methods for estimating $\tau$: restricted maximum likelihood (REML) and a nearest-neighbors estimation approach.

\noindent
\textit{REML Estimation.}  REML estimation is widely used for estimating variance components in linear mixed models \citep{jiang2021} and in small area estimation \citep{rao2015}. The standard REML approach maximizes the likelihood associated with $\mathbf{A}^{T}\mathbf{Y}$, where the $n \times (n-p)$ matrix $\mathbf{A}$ satisfies $\mathbf{A}^{T}\mathbf{X} = \mathbf{0}$, and under working model (\ref{eq:working_model}), the standard REML objective function \citep{harville1974} to minimize reduces to 
\begin{equation}
\ell_{REML}(\tau) = \log\det( \mathbf{W}_{\tau} ) + \log\det( \mathbf{X}^{T}\mathbf{W}_{\tau}^{-1}\mathbf{X}) + \mathbf{Y}^{T}\mathbf{P}_{\tau}\mathbf{Y}, 
\label{eq:reml_objective}
\end{equation}
where $\mathbf{W}_{\tau} = \textrm{diag}\{ \tau^{2} + \sigma_{1}^{2}, \ldots, \tau^{2} + \sigma_{K}^{2} \}$ and
$\mathbf{P}_{\tau} = \mathbf{W}_{\tau}^{-1} - \mathbf{W}_{\tau}^{-1}\mathbf{X}(\mathbf{X}^{T}\mathbf{W}_{\tau}^{-1}\mathbf{X})^{-1}\mathbf{X}^{T}\mathbf{W}_{\tau}^{-1}$.

\noindent
\textit{Nearest Neighbors Estimation.}
While REML estimation of $\tau^{2}$ is an effective approach, the REML objective function (\ref{eq:reml_objective}) is based on working model assumption (\ref{eq:working_model}), and alternative methods that do not make the assumption that $\mu_{k} = \mathbf{x}_{k}^{T}\bbeta$ can provide a more robust approach to estimating $\tau^{2}$. One such method is the nearest-neighbor residual variance estimator detailed in \cite{Devroye2018}. Their estimator is designed to estimate the term $\min_{g} E\big\{ \big( g(X_{i}) - Y_{i} \big)^{2}  \big\}$, when one has observed random variable pairs $(X_{1}, Y_{1}), \ldots, (X_{n}, Y_{n})$. 

In our setting, where the distribution of $Y_{k}$ is determined by model (\ref{eq:true_model}), we expect that $\min_{g} E\{ \big( g(\mathbf{x}_{k}) - Y_{k} \big)^{2}  \}$ will be close to $\tau^{2} + \sigma_{k}^{2}$, and hence, we need to adjust for the values of $\sigma_{k}^{2}$ when applying the method of \cite{Devroye2018}. When including an adjustment for $\sigma_{1}^2, \ldots, \sigma_{K
}^{2}$, the one-nearest neighbor estimator of $\tau^{2}$ is given by
\begin{equation}
\hat{\tau}^{2}_{1-NN} = \frac{1}{K}\sum_{k=1}^{K} Y_{k}^{2} - \frac{1}{K_{T}}\sum_{k \in \mathcal{D}} Y_{k}Y_{k}' - \frac{1}{K_{T}}\sum_{k \in \mathcal{D}} \sigma_{k}^{2},
\label{eq:nn_tau_est}
\end{equation}
where $\mathcal{D}$ is a set of indices denoting a test set of observations, 
$K_{T}$ is the number of elements in $\mathcal{D}$, and $Y_{k}'$ is the value of $Y$ associated with the nearest neighbor of $\mathbf{x}_{k}$ in the training set.
That is, $Y_{k}' = Y_{j^{*}}$, where $j^{*} = \arg\min_{j \not\in \mathcal{D}} [(\mathbf{x}_{j} - \mathbf{x}_{k})^{T}(\mathbf{x}_{j} - \mathbf{x}_{k})]$.
Following the recommendation in \cite{Devroye2018}, we set the size of the test set $\mathcal{D}$ to include approximately half of the units. Specifically, we set the size of $\mathcal{D}$ to be $K/2$ if $K$ is even and to be $(K + 1)/2$ if $K$ is odd. One should also note that it is possible for $\hat{\tau}^{2}_{1-NN}$ in (\ref{eq:nn_tau_est}) to be negative. When this occurs, we set the estimate of $\tau^{2}$ to be the REML estimate.

\subsection{Targeted Estimation for Proper Percentiles}
\label{sec:uniform_constraint}

The collection of posterior expected ranks or posterior expected percentiles
will not, in general, constitute a set of ``proper'' rankings or a set of ``proper'' percentiles.
That is, the posterior expected ranks will typically not generate a set of integers that exhaust the integers from $1$ to $K$, and similarly, the posterior expected percentiles will not exhaust
the values $\tfrac{1}{K+1}, \ldots, \tfrac{K}{K+1}$. Rather, the posterior expected percentiles 
will typically be a collection of numbers that appear to be shrunken towards the mid-percentile $1/2$ and will not fully populate the values of $\tfrac{1}{K+1}, \ldots, \tfrac{K}{K+1}$.

In many contexts, the task of interest is to generate a proper ranking (or, equivalently, proper percentiling) of the units, where one reports the unit that is ranked first, the unit that is ranked second, etc. As reported in \cite{Lin2006}, when targeting rank squared-error loss, the optimal ranking
procedure subject to the constraint that one reports a proper ranking that exhausts the integers from $1$ to $K$ is obtained by ranking the posterior expected ranks. Applying this result to our setting would imply that the optimal way, for a fixed $\bbeta$, to generate percentiles, subject to the constraint that the percentiles constitute a proper set of percentiles, is to use the percentiles of the posterior expected population percentiles. However, developing an unbiased risk estimator similar to (\ref{eq:Qhat}) that incorporates this constraint leads to an objective function that can be computationally challenging to optimize and does not notably improve performance compared to directly using the percentiles of ROPPER. As an alternative, we propose a more direct approach that ensures the reported percentiles are as close to the ROPPER values $R_{k}(\hatbbeta_{r}, \tau, Y_{k})$ as possible. This is done by simply computing the percentiles of $R_{k}(\hatbbeta_{r}, \tau, Y_{k})$, as summarized in the proposition below.
\begin{prop}
The collection of integers $A_{1}, \ldots, A_{K}$ that minimize \\ $\sum_{k=1}^{K} \big\{ \tfrac{1}{K+1}A_{k} - R_{k}(\hatbbeta_{r}, \tau, Y_{k}) \big\}^{2}$, subject to the constraint that $A_{1}, \ldots, A_{K}$ exhaust the integers $1, \ldots, K$ is given by
$A_{k} = \textrm{rank}\big( R_{k}(\hatbbeta_{r}, \tau, Y_{k}) \big)$.
\end{prop}

\subsection{Optimization}

A major computational challenge that arises when minimizing an order-H approximation $\hat{Q}_{\tau}^{(H)}(\bbeta)$ of the unbiased risk estimator (\ref{eq:Qhat}) is that this objective function is nonconvex. While we have found that ``general-purpose'' optimizers based on quasi-Newton methods perform well in many examples, we have observed that, in a few cases, these optimizers have unstable performance and do not converge after many iterations.

To minimize the first-order approximation $\hat{Q}_{\tau}^{(1)}(\bbeta)$ of the unbiased risk estimator, we have developed a majorize-minimization (MM) algorithm \citep{lange2016} that is guaranteed to decrease the value of $\hat{Q}_{\tau}^{(1)}(\bbeta)$ in every 
iteration. MM algorithms are generally more numerically stable and less sensitive to the choice of initial value. A main advantage of using an MM algorithm for minimization of $\hat{Q}^{(1)}(\bbeta)$ is that, if the initial value is set to the MLE $\hatbbeta_{MLE}$, then we are guaranteed to achieve
a smaller value of $\hat{Q}^{(1)}$ compared to the MLE. That is, 
we can ensure that $\hat{Q}_{\tau}^{(1)}(\bbeta^{*}) \leq \hat{Q}_{\tau}^{(1)}(\hatbbeta_{MLE})$, 
where $\bbeta^{*}$ is the parameter vector obtained after running our MM algorithm to convergence.

To develop our MM algorithm, we identified a majorizing quadratic function for $\hat{Q}^{(1)}(\bbeta)$ that leads to an update of the parameter vector found by solving a weighted least-squares problem. Specifically, in the $t^{th}$ iteration of the MM algorithm, the parameter vector update can be expressed as 
\begin{equation}
\bbeta^{(t+1)} = \Bigg(\mathbf{X}^{T}\mathbf{V}^{T}(\mathbf{W}_{1}^{(t)} + \tfrac{1}{3}\mathbf{W}_{2}^{(t)})\mathbf{V}\mathbf{X}\Bigg)^{-1}\Big( \mathbf{X}^{T}\mathbf{V}^{T}\big[ \mathbf{W}_{1}^{(t)}\mathbf{V}\mathbf{Y}  
+ \mathbf{W}_{2}^{(t)}\{ \mathbf{d}(\bbeta^{(t)})
+ \tfrac{1}{3}\mathbf{V}\mathbf{X}\bbeta^{(t)} \} \big] \Big),
\label{eq:mm_alg_update}
\end{equation}
where $\mathbf{V} = \textrm{diag}\{ V_{1,\tau}, \ldots, V_{K,\tau}\}$ and $d(\bbeta^{(t)})$ is the $K \times 1$ vector whose $k^{th}$ element is 
\begin{equation}
d_{k}(\bbeta^{(t)}) = \frac{2\phi(V_{k,\tau}(Y_{k} - \mathbf{x}_{k}^{T}\bbeta^{(t)}))D( V_{k,\tau}(Y_{k} - \mathbf{x}_{k}^{T}\bbeta^{(t)}) ) }{ 1 - D^{2}( V_{k,\tau}(Y_{k} - \mathbf{x}_{k}^{T}\bbeta^{(t)}) ) }, \nonumber
\end{equation}
where $D$ is the function defined as $D(u) = \Phi(u) - 1/2$ and where $\mathbf{W}_{1}^{(t)}$ and $\mathbf{W}_{2}^{(t)}$ are
$K \times K$ diagonal matrices with diagonal elements $w_{k1}(\bbeta^{(t)})$ and $w_{k2}(\bbeta^{(t)})$ respectively.  
These diagonal elements are defined as
\begin{eqnarray}
w_{k1}(\bbeta^{(t)}) &=& C_{t} V_{k,\tau}\phi(V_{k,\tau}(Y_{k} - \mathbf{x}_{k}^{T}\bbeta^{(t)})) \nonumber\\
w_{k2}(\bbeta^{(t)}) &=& \frac{C_{t}\sqrt{\pi} }{\tau\sqrt{2}}\Big\{  1 - D^{2}( V_{k,\tau}(Y_{k} - \mathbf{x}_{k}^{T}\bbeta^{(t)}) ) \Big\}, \nonumber
\end{eqnarray}
where $C_{t} > 0$ is a constant chosen to ensure that $\sum_{k=1}^{K} \{ w_{k1}(\bbeta^{(t)}) + w_{k2}(\bbeta^{(t)})\} = 1$.

The descent property of the iterative scheme (\ref{eq:mm_alg_update}) is stated 
formally by the theorem below and is proved in Appendix D.
\vspace{-0.2cm}

\begin{theorem}
For a fixed value of $\tau$, iterates $\bbeta^{(0)}, \bbeta^{(1)}, \bbeta^{(2)}, ...$ from the updating scheme defined in (\ref{eq:mm_alg_update}) are guaranteed to decrease the first-order approximation of the unbiased risk estimate $\hat{Q}_{\tau}(\bbeta)$ in every iteration. That is, for any integer $t \geq 0$,
\begin{equation}
\hat{Q}_{\tau}^{(1)}(\bbeta^{(t+1)}) \leq \hat{Q}_{\tau}^{(1)}(\bbeta^{(t)}). \nonumber
\end{equation}
\end{theorem}

\section{Simulation Studies}
\subsection{Setting with an Unmodeled Latent Subgroup}
\label{sec:basic}
\vspace{-0.3cm}

We assessed the properties of ROPPER empirically through three simulation studies. In all studies, 
we used ROPPER with the order-1 approximation $\hat{Q}^{(1)}(\bbeta)$ of the unbiased risk estimator (\ref{eq:Qhat}). Appendix E contains evaluations of ROPPER for different choices of the approximation order $H$.

In the first setting, we generated the unit-level summary statistics $Y_k$ from the following model:
\begin{equation}
Y_k = \beta_0 + \beta_1 X_k + \varv_k + e_k, \nonumber 
\end{equation}
where $X_k \sim \textrm{Bernoulli}(0.5)$, $\varv_k \sim N(0,\tau^2)$, and $e_k \sim N(0,\sigma^2/n_k)$. The parameters were set to $\beta_0=1$ and $\tau^2=1$, with $\beta_1$ varying from -1 to 1. We considered the following choices of $K$ and $\sigma^{2}$: $K \in \{50, 500 \}$, and $\sigma^2 \in \{2, 5\}$. Using this simulated data, we estimated $\beta_0$ from an underspecified intercept-only model using both our proposed RFURE approach ($\hat{\beta}_{0,r}$) and the conventional MLE ($\hat{\beta}_{0,MLE}$) for comparison. Here, we used an underspecified model to highlight realistic scenarios in which the proposed ranking-targeted estimator and the MLE are expected to be meaningfully different. 

\begin{figure}[h!]
\centering
\includegraphics[width=0.9\textwidth]{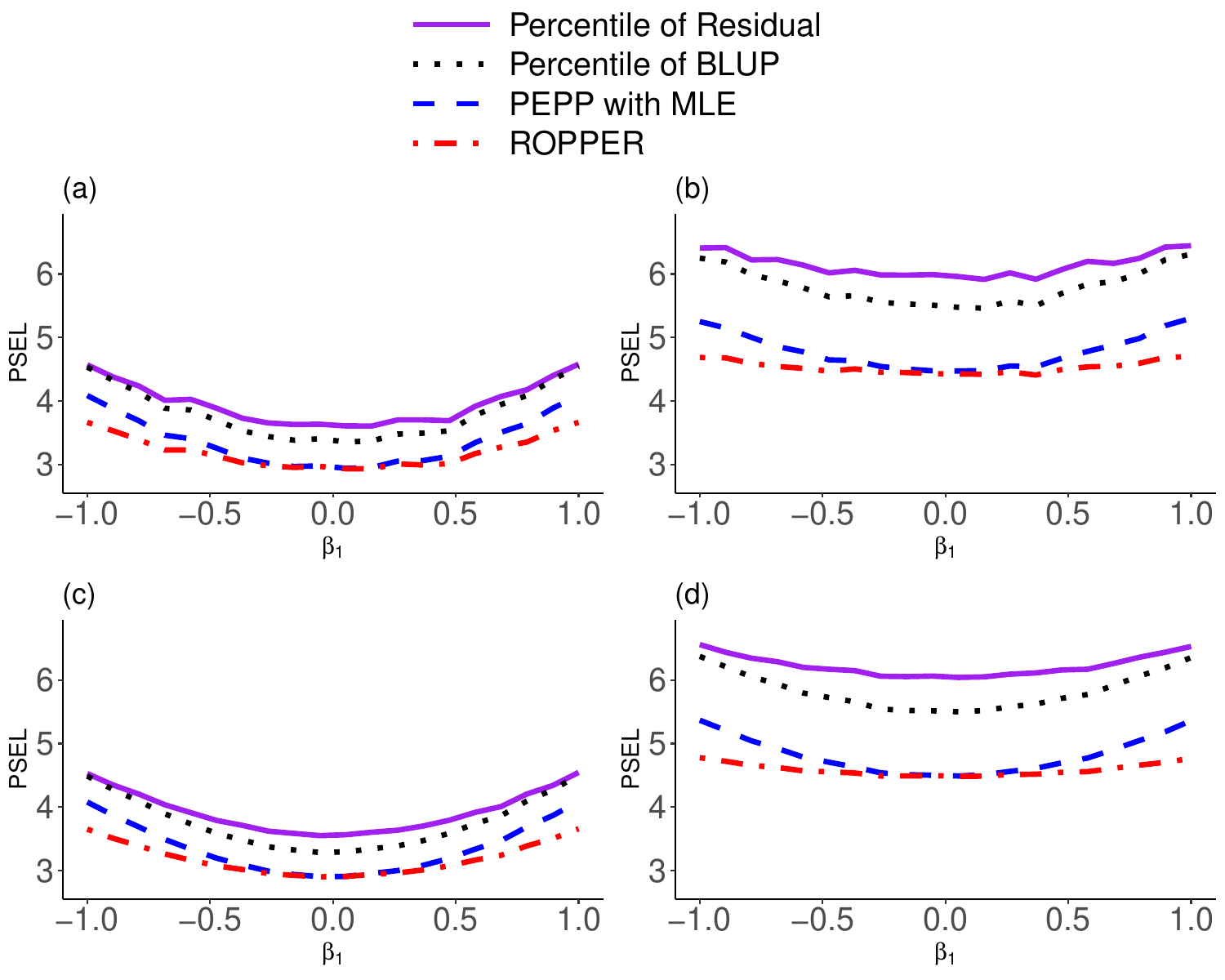}
\caption{Percentile squared-error loss (PSEL) for different ranking methods in the unmodeled latent subgroup simulation setting. Panels: (a) $K = 50$ and $\sigma^{2} = 2$
(b) $K = 50$ and $\sigma^{2} = 5$
(c) $K = 500$ and $\sigma^{2} = 2$
(d) $K = 500$ and $\sigma^{2} = 5$, where $\beta_1$: effect size of unmeasured covariate, $K$: number of units, $\sigma^2$: residual variance. PEPP: Posterior Expected Population Percentile. ROPPER: Ranking-Optimized Population Posterior Expected peRcentiles. PSEL values are based on 1000 simulation iterations.}
\label{fig:latentcluster_simresults}
\end{figure}

We calculated the PEPP, $R_{k}(\hat{\beta}_{0,r}, \tau, Y_{k})$ and $R_{k}(\hat{\beta}_{0,MLE}, \tau, Y_{k})$, according to (\ref{eq:pep_definition}), and additionally, we computed the percentiles of the residuals, defined as $(Y_k - \hat{\beta}_{0,MLE})$, and the BLUP of $\varv_k$, defined as $B_{k,\tau}(Y_k - \hat{\beta}_{0,MLE})$. For each of these four percentiling methods, we calculated the PSEL loss function by plugging in $R_{k}(\hat{\beta}_{0,r}, \tau, Y_{k})$, $R_{k}(\hat{\beta}_{0,MLE}, \tau, Y_{k})$, $\textrm{perc}(Y_k-\hat{\beta}_{0,MLE})$, or $\textrm{perc}(B_{k,\tau}(Y_k-\hat{\beta}_{0,MLE}))$ as the $A_k$ in (\ref{eq:psel_def}). We repeated the calculations described above for 1000 simulated datasets, and averaged the PSEL results over all iterations. 

As shown in Figure \ref{fig:latentcluster_simresults}, we found that ROPPER always had the lowest PSEL, followed by the PEPP using $\hat{\beta}_{0,MLE}$. The residual and BLUP percentiles consistently performed worse than the PEPP with $\hat{\beta}_{0,MLE}$, and the BLUP percentiles almost always achieved a modestly smaller value of the PSEL than that of the residual percentiles. These differences increased with the magnitude of the absolute value of $\beta_1$ (the effect of the unmeasured covariate $X_k$), suggesting that ROPPER can be more robust to this form of model misspecification. These findings provide empirical evidence that the proposed ranking-targeted estimation approach offers superior ranking performance compared to using the MLE and that using the PEPP to define the ranking function provides additional advantages over ranking through use of the BLUP.

\subsection{A Nonlinear Regression Function with Four Covariates}
\label{sec:five}

In our second simulation setting, we introduced more complexity into the fixed effects component of the model used to generate $Y_k$ by considering a nonlinear fixed effects function that includes four covariates. Specifically, the $Y_{k}$ were generated from the following model:
\begin{equation*}
\begin{aligned}
    Y_{k} &= \beta_{0} + \beta_{1}X_{k1} + \beta_{2}X_{k2}^{\gamma_{1}}(1 - X_{k1})^{\gamma_{2}}+ \beta_{3}X_{k3}^{\gamma_{3}}(1 - X_{k1})^{\gamma_{4}} + \\ &\beta_{4}X_{k1}^{\gamma_{5}}(1 - X_{k2})^{\gamma_{6}}(1 - X_{k3})^{\gamma_{7}} + \beta_{5}X_{k4} + \varv_{k} + e_{k},
\end{aligned}
\end{equation*}

\noindent where $X_{kj} \sim\textrm{Uniform}(0, 1), j = 1, \ldots, 5$, with an initial design of $\gamma_1=1$, $\gamma_2=2$, $\gamma_3=0.5$, $\gamma_4=1.5$,$\gamma_5=-1$, $\gamma_6=0.75$, $\gamma_7=3$, $\beta_0=-1$, $\beta_1=1$, $\beta_2=0.5$, $\beta_3=0$, $\beta_4=-0.5$, and $\beta_6$ varied from -1 to 1. In a second design, we multiplied all $\gamma$ values by 1.5 to increase the level of model misspecification. The random components $\varv_k$ and $e_k$ had the same distributions as in Section \ref{sec:basic}. 

Using the simulated data, we fit a misspecified linear model where the fixed effects means were assumed to have the form $\mu_{k} = \beta_{0} + \sum_{j=1}^{4} X_{kj}\beta_{j}$ and calculated the PSEL values for each ranking method described in Section \ref{sec:basic} using 1000 iterations. To additionally assess ROPPER under the restriction that one must use proper percentiles, we also evaluated the ranking performance of using the empirical percentiles of ROPPER and the empirical percentiles of the PEPP with the MLE. 

In this simulation study, we also explored the impact of two common sources of model misspecification that can notably affect the performance of the traditional MLE approach. First, we allowed $\varv_k$ and $X_{kj}$ to be correlated by drawing $\varv_k \sim N\big( \alpha_1 (\boldsymbol{X}_k^{\top}\boldsymbol{\beta}-\frac{1}{K}\sum_{k=1}^K\boldsymbol{X}^{\top}\boldsymbol{\beta}),\tau^2 \big)$, and we varied $\alpha_1$ to study the impact of the magnitude of this correlation. Secondly, we allowed $\varv_k$ to be correlated with the unit size $n_k$ by simulating $\varv_k \sim N(\alpha_2 (n_k-\frac{1}{K}\sum_{k=1}^Kn_k),\tau^2)$, and varied $\alpha_2$ to study the impact of the magnitude of this correlation.

\begin{figure}[h!]
\centering
\includegraphics[width=\textwidth]{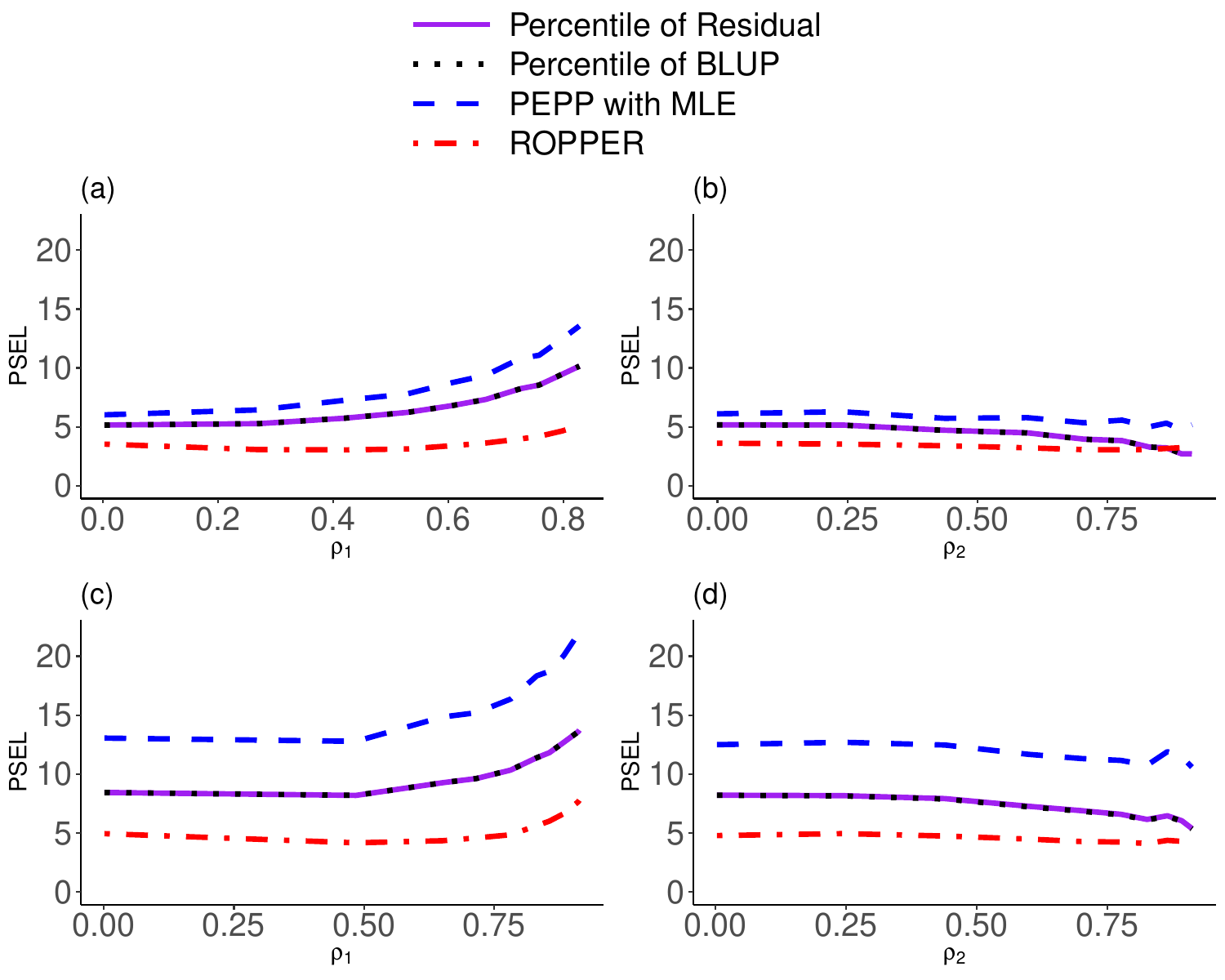}
\caption{Simulation setting with a nonlinear regression function based on four covariates. Percentile squared-error loss (PSEL) for different ranking methods, where $\rho_1=Cor(v_k, \boldsymbol{X}_k^{\top}\boldsymbol{\beta})$ and $\rho_2=Cor(v_k, n_k)$. Panels (a) and (b) have $\gamma_1=1$, $\gamma_2=2$, $\gamma_3=0.5$, $\gamma_4=1.5$,$\gamma_5=-1$, $\gamma_6=0.75$, $\gamma_7=3$ whereas panels (c) and (d) have all $\gamma$ values multiplied by 1.5 to introduce more model misspecification. PEPP: Posterior Expected Population Percentile. ROPPER: Ranking-Optimized Population Posterior Expected peRcentiles. Based on 1000 simulation iterations.}
\label{fig:setting2}
\end{figure}

\begin{figure}[h!]
\centering
\includegraphics[width=0.9\textwidth]{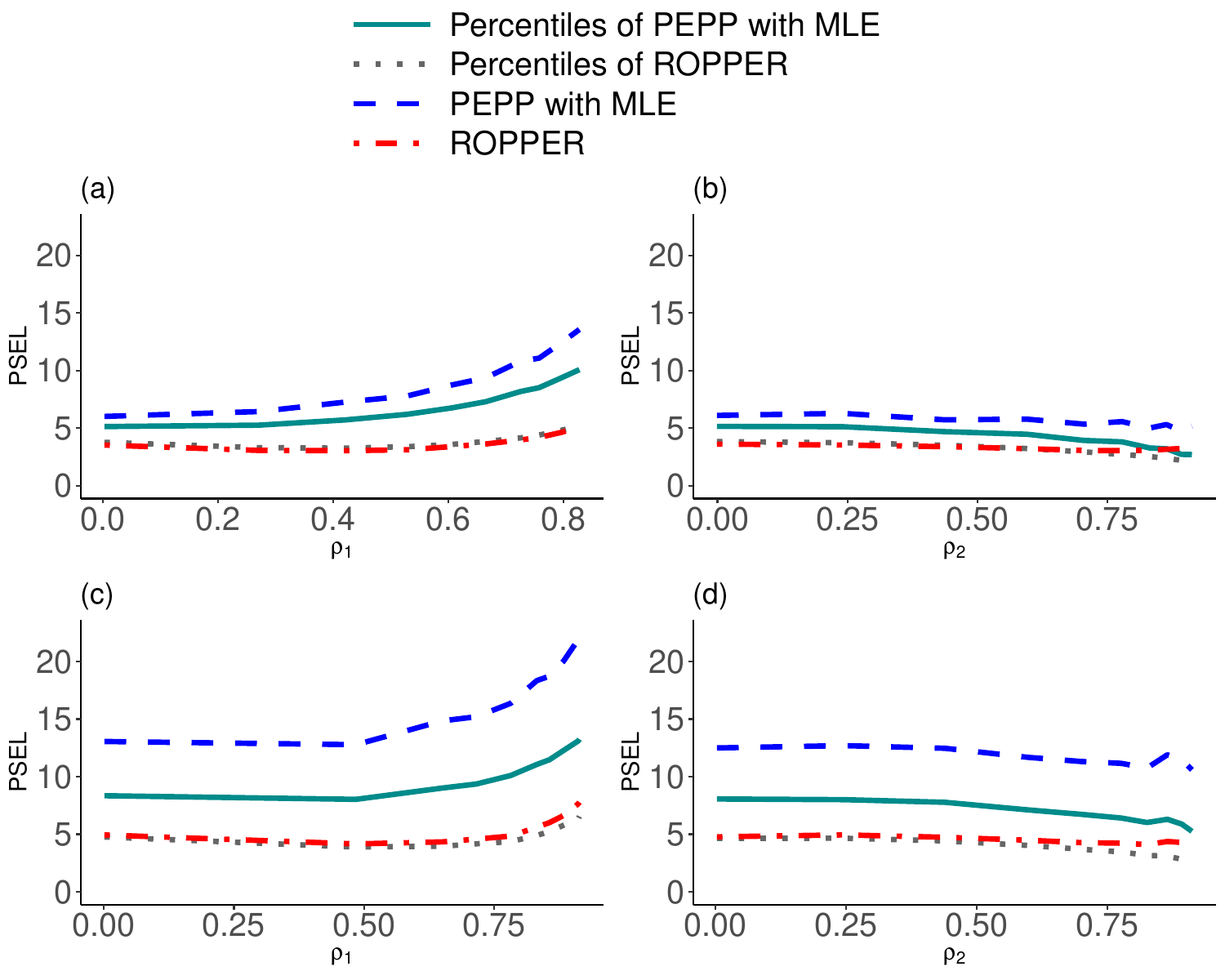}
\caption{Simulation setting with a nonlinear regression function based on four covariates. Percentile squared-error loss (PSEL) for different ranking methods, where $\rho_1=Cor(v_k, \boldsymbol{X}_k^{\top}\boldsymbol{\beta})$ and $\rho_2=Cor(v_k, n_k)$. Panels (a) and (b) have $\gamma_1=1$, $\gamma_2=2$, $\gamma_3=0.5$, $\gamma_4=1.5$,$\gamma_5=-1$, $\gamma_6=0.75$, $\gamma_7=3$ whereas panels (c) and (d) have all $\gamma$ values multiplied by 1.5 to introduce more model misspecification. PEPP: Posterior Expected Population Percentile. ROPPER: Ranking-Optimized Population Posterior Expected peRcentiles. PSEL values are based on 1000 simulation iterations.}
\label{fig:supplement_simresults}
\end{figure}

As shown in Figure \ref{fig:setting2}, we observed a very similar pattern of performance across methods in this setting as observed in Section \ref{sec:basic}. In addition, taking the percentiles of the PEPP with MLE reduced the PSEL values compared to using the original PEPP values (Figure \ref{fig:supplement_simresults}). Using the percentiles of ROPPER rather than the original ROPPER values had very little impact on the PSEL in this simulation setting. From the results presented in both Figures \ref{fig:setting2} and \ref{fig:supplement_simresults}, we also found that the PSEL values tended to increase strongly with the correlation between $\varv_k$ and the $X_{kj}$, but not with the correlation between $n_k$ and $\varv_k$, which is consistent with the findings of previous literature \citep{neuhaus2011,kalbfleisch2013}.

\subsection{Simulation Emulating Real-world Education Data}
\label{sec:education_emulate}

In our third simulation setting, we generated data that closely resembled our motivating application of elementary school rankings, as described in Section \ref{sec:application}. Specifically, we used the actual covariate values and corresponding fixed effects regression coefficient estimates from our real data analysis to simulate outcomes $Y_k$ for 862 schools (i.e., the number of schools in our real dataset). 

\begin{figure}[h!]
\centering
\includegraphics[width=\textwidth]{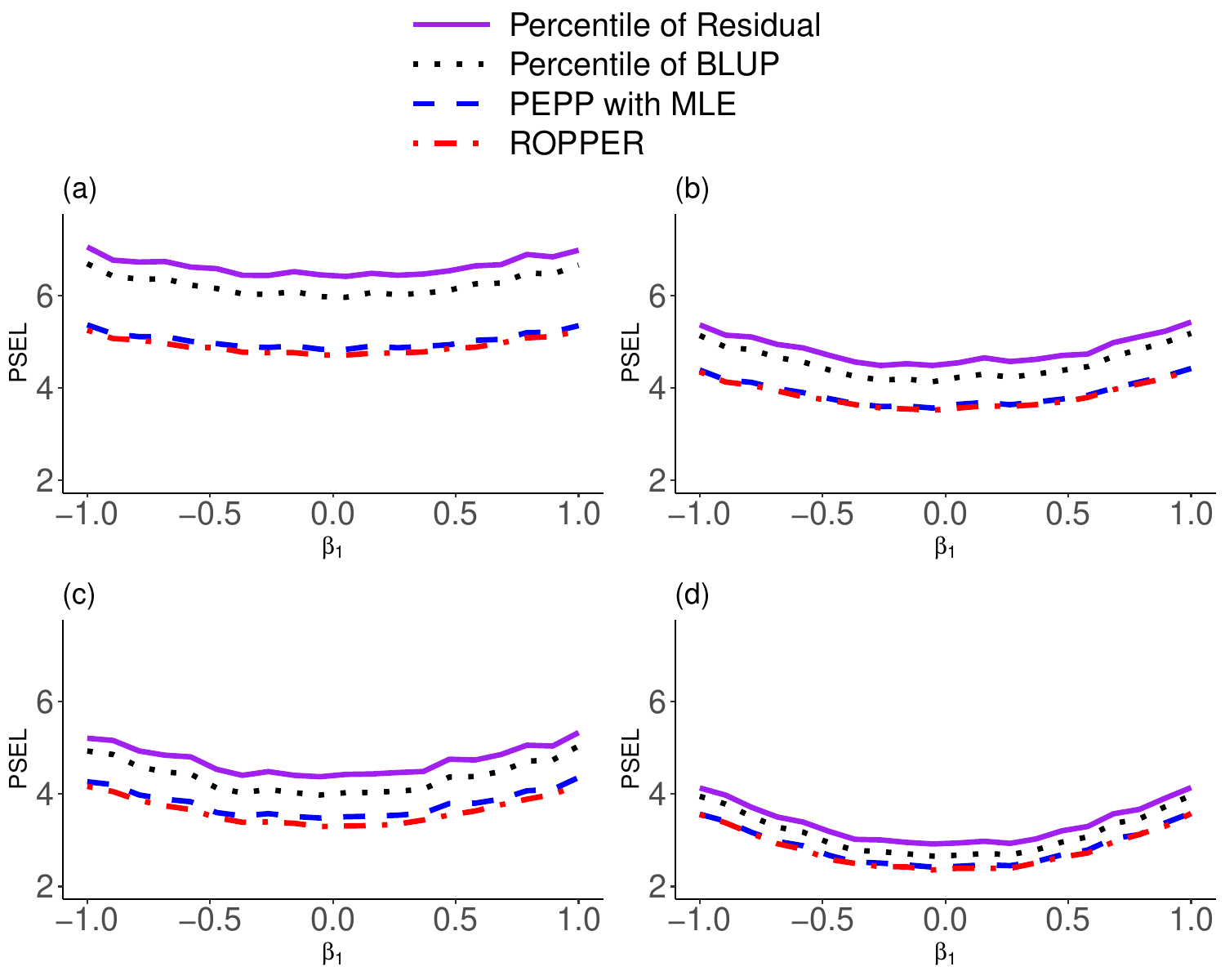}
\caption{Simulation setting based on emulating the real-world education data. Percentile squared-error loss (PSEL) for different ranking methods. Panels: (a) $\max(n_k)=20$ and $\sigma^{2} = 10$
(b) $\max(n_k)=20$ and $\sigma^{2} = 5$
(c) $\max(n_k)=50$ and $\sigma^{2} = 10$
(d) $\max(n_k)=50$ and $\sigma^{2} = 5$, where $\beta_1$: effect size of unmeasured covariate, $\sigma^2$: residual variance, $n_k$: $k$th unit size. PEPP: Posterior Expected Population Percentile. ROPPER: Ranking-Optimized Population Posterior Expected peRcentiles. PSEL values are based on 1000 simulation iterations.}
\label{fig:setting3_simresults}
\end{figure}

In the data generation process, we also incorporated two realistic sources of model misspecification that are plausible in a school ranking analysis. First, given that our motivating education dataset was collected from an observational study, there is always the potential for unmeasured confounding. Therefore, we included a latent binary covariate in the data generating model described above, with an associated regression coefficient $\beta_1$ that we varied from -1 to 1. This latent covariate was excluded when performing estimation of $\hatbbeta_{r}$ and $\hatbbeta_{MLE}$. Second, it is likely that some of the covariates we explore in our education example (e.g., average household income) may be correlated with the random effects $\varv_k$ representing underlying school quality. Thus, we induced correlation between $\varv_k$ and the $X_{kj}$ values using a method similar to that in Section \ref{sec:five}. In these simulations, we let $\tau^2=1$ and set $\sigma^2$ equal to either 2 or 5. We generated the school sizes from a discrete uniform distribution as $n_k \sim \textrm{Unif}\{1,n_{\textrm{max}}\}$, with $n_{\textrm{max}}$ set to either 20 or 50. Figure \ref{fig:setting3_simresults} shows again that the ROPPER approach outperformed each of the MLE-based ranking methods across all values of $\beta_{1}$, and the magnitude of these differences was sensitive to the distribution of $n_{k}$.

\subsection{Additional Simulation Studies}

Through several further simulation studies, we explored additional properties of the proposed methods (Appendix E). In the first study, we replicated the simulation design from the unmodeled latent subgroup study but with $\varv_k$ simulated from various non-normal distributions (holding the variance fixed at $\tau^{2}$). We observed in Figure \ref{fig:other_distribution_simresults} that the patterns in PSEL for different random effect distributions were very similar to the original results in Figure \ref{fig:latentcluster_simresults}. However, the method using the residual percentiles was more sensitive to departures from normality. In a separate simulation, we compared the results of the REML with the one-nearest neighbors (KNN) $\tau$-estimation approaches using the simulation setting of Section \ref{sec:education_emulate} and observed that these methods had substantial agreement (Figure \ref{fig:knn_compare}).
\vspace{-0.3cm}

\section{Application: Education Data}
\label{sec:application}
\vspace{-0.3cm}

We applied our proposed methodology to data from the Early Childhood Longitudinal Study (ECLS), which collected a wide range of information on elementary school students through surveys and administered exams \citep{ECLS}. In this analysis, our objective was to rank the 862 schools represented in this study based on the underlying quality of their mathematics education, as represented by the random effects $\varv_k$. To do this, we defined the unit-level outcome variables $Y_k$ as the average mathematics exam score in each school, and we adjusted for the following fixed covariates $\mathbf{x}_k$: school locale (city, suburban, town, rural), school type (public or private), average student family income, proportion of teachers with graduate education, and proportion of students having a parent with a college education \citep{tourangeau2015early}. We implemented the proposed ROPPER procedure, with $\tau$ estimated using REML. In addition, we used this estimate of $\tau$ together with the MLE of the fixed effects regression coefficients to construct percentiles based on residual ranking, PEPP with MLE, and ranking the components of the BLUP. 

\begin{table}[h!]
    \caption{Comparison of coefficient estimates for the association between school-level covariates and student math performance, using maximum likelihood estimation ($\hatbbeta_{MLE}$) versus the proposed RFURE ($\hatbbeta_r$). Coefficients estimates were computed using average mathematics scores from the Early Childhood Longitudinal Study.} 
    \begin{tabular}{lcc}
    \hline
        Variable & $\hatbbeta_{MLE}$ & $\hatbbeta_r$ \\
    \hline
  Intercept & \llap{-}0.895 & \llap{-}0.866 \\ 
  Suburban Locale & \llap{-}0.047 & \llap{-}0.007 \\ 
  Town Locale  & \llap{-}0.001 &  0.025 \\ 
  Rural Locale &  0.000 &  0.035 \\ 
  Private School & \llap{-}0.063 &  0.045 \\ 
  Average Household Income &  0.031 &  0.026 \\ 
  Proportion of Teachers with Graduate Degree  & \llap{-}0.073 & \llap{-}0.025 \\ 
  Proportion of First Parents with College Degree &  0.329 &  0.279 \\ 
  Proportion of Second Parents with College Degree &  0.114 &  0.158 \\ 
  \hline
    \end{tabular}
    \label{tab:est_educ}
\end{table}

\begin{figure}[h!]
\centering
\includegraphics[width=\textwidth]{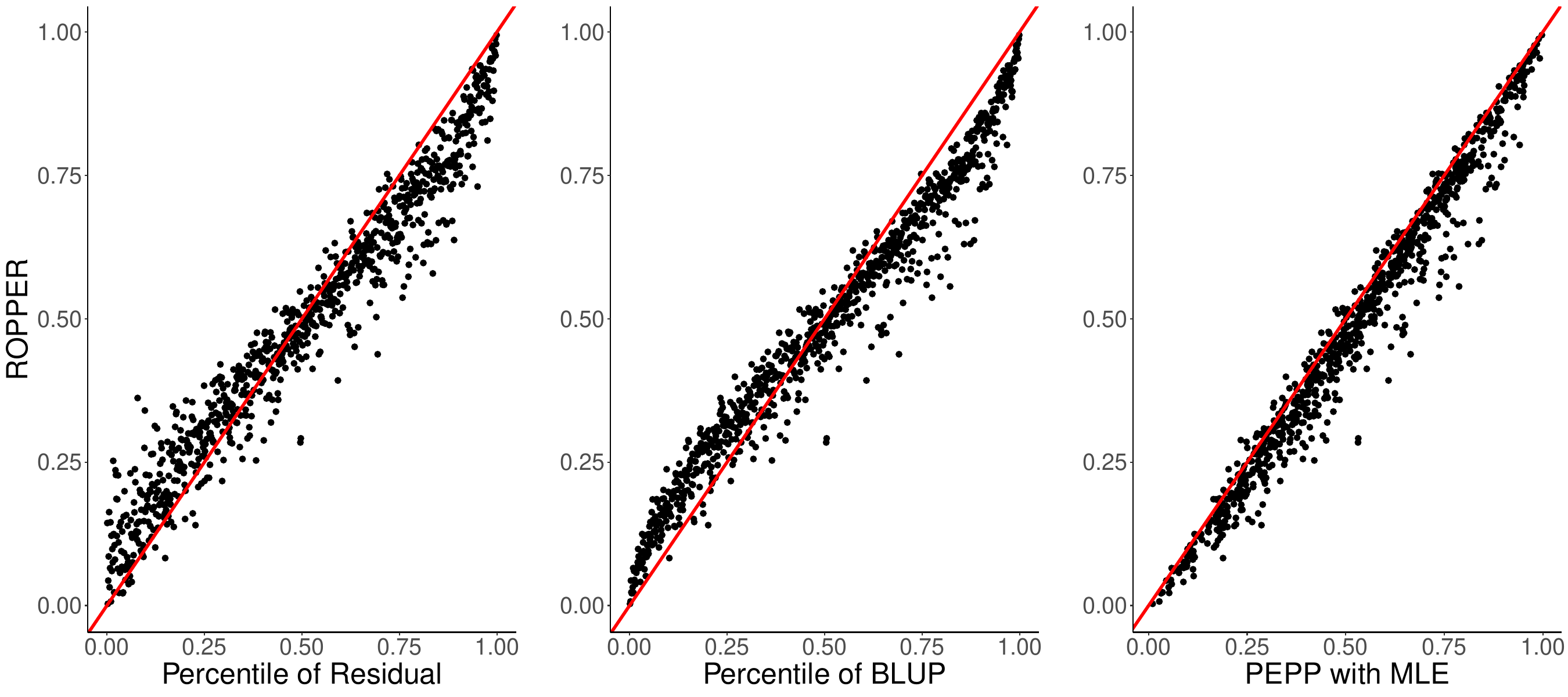}
\caption{Comparison of the math education percentiles of 862 schools from the Early Childhood Longitudinal Study, based on the proposed ROPPER (Ranking-Optimized Population Posterior Expected peRcentiles) versus the percentile of the residual, the percentile of the Best Linear Unbiased Predictor (BLUP), and the posterior expected population percentile (PEPP) with the maximum likelihood estimator (MLE). Red diagonal line: intercept of zero and slope of one.}
\label{fig:real_comparison}
\end{figure}

As reported in Table \ref{tab:est_educ}, we found that there were nontrivial differences between the components of the RFURE $\hatbbeta_r$ and the MLE $\hatbbeta_{MLE}$. When evaluating the actual school rankings, we also observed meaningful variation in the school rankings across methods, which can be seen in Figure \ref{fig:real_comparison}. More specifically, when applying ROPPER, 105, 78, and 68 schools moved down a decile or more compared to their ranking based on the percentiles of the residual, percentiles of the BLUP, and PEPP with MLE, respectively. In addition, with ROPPER, 60 and 29 schools moved up a decile compared to their rankings based on the percentiles of the residuals and BLUP, respectively. These results demonstrate that using an estimation approach tailored to ranking goals can provide meaningfully different results compared to more conventional estimation strategies. 

\vspace{-0.5cm}

\section{Discussion}
\label{s:discussion}
In this paper, we exclusively focused on the problem of ranking unit-level effects after adjusting for unit-level covariates. We focused on this due to the interest, in many applied contexts, in characterizing unit-level performance beyond the impact of commonly collected covariates. However, in many contexts, a more natural goal is to directly rank the unit-level conditional means $\mu_{k} + \varv_{k}$ without considering the role of covariates in explaining the variation in the unit-level means. Our ROPPER framework could potentially be modified for use in such settings. However, because the percentiles of the unadjusted unit-level conditional means $\mu_{k} + \varv_{k}$ would depend on the unknown unit-level fixed mean parameters $\mu_{k}$, the main challenge in adapting our approach would be choosing how to appropriately represent the population percentiles. One possible choice for defining the population percentiles would be to assume that the unit-level fixed means $\mu_{k}$ also represented draws from a shared distribution and to model this unknown distribution non-parametrically. If one could derive a tractable form for the posterior expected population percentiles using this approach, it would facilitate the development of an unbiased risk estimator similar to the one developed in this paper. While ranking $\mu_{k} + \varv_{k}$ rather than $\varv_{k}$ is an important goal that could arise in many contexts, we leave this to future work due to the substantial number of additional challenges in adapting our approach to effectively incorporate this ranking objective.

The assumption of working model (\ref{eq:working_model}) that the unit-level fixed means $\mu_{k}$ are linear combinations of the observed covariates was used to derive the form of the PEPP $R_{k}(\bbeta, \tau, Y_{k})$, but this linearity assumption is not needed for the unbiased property of the unbiased risk estimator (\ref{eq:Qhat}) to hold. In this sense, the RFURE $\hatbbeta_{r}$ of the regression coefficients is robust because it utilizes an estimate of risk that does not rely on the linearity assumption. However, the unbiased property of the risk estimator (\ref{eq:Qhat}) does rely on the normality assumption of the random effects. Although our approach does seem reasonably robust to modest departures from the normality assumption of the random effects distribution (see Appendix E), there are many settings where the random effects would be better modeled using distributions with features that are quite distinct from those of Gaussian distributions. For example, it is commonly noted that the underlying effects of interest often have a distribution that is quite heavy-tailed (e.g., \cite{ohlssen2007}) or have a distribution with multiple modes. One potential way to extend our approach to better accommodate such scenarios would be to assume that the random effects arise from a discrete mixture of normal distributions (e.g., \cite{stephens2017}) rather than a single Gaussian distribution. When conditioning on a particular mixture label, the form of the PEPP would remain unchanged, and hence, the PEPP would take the form of a weighted average of the mixture-component conditional posterior expected population percentiles. To fully develop this approach, one would still need to appropriately modify the unbiased risk estimator, but the Gaussian mixture form of the posterior expected population percentiles could provide a useful class of percentiles to consider.

Our estimation procedure was motivated by the goal of minimizing the expected percentile squared-error loss, where units have equal weight in the loss function. However, our procedure could be easily modified to handle a weighted percentile squared-error loss if the unit-level weights were fixed and pre-specified. To do this, one would only need to estimate the fixed effects regression coefficients by minimizing a weighted version of our unbiased risk estimator. However, in many cases, one would want the weights to reflect the magnitude of the underlying random effects in some way, and pre-specifying unit-specific weights to reflect this would be more challenging.
For example, one may want to employ a weight function that more heavily weights ``extreme'' values of the random effects \citep{mcculloch2023}. Developing a procedure that incorporates such a random effects-dependent weight function would be less straightforward and would require deriving a new unbiased risk estimator. Nevertheless, for simple choices of a weight function, one could potentially derive an unbiased risk estimator using techniques similar to those we used to derive the unbiased risk estimator developed in this paper.

\medskip
\begin{center}
{\large\bf Supplementary Material}
\end{center}
\vspace{-.6cm}

\begin{description}
\item[R-package:] An R-package \verb"ranefranking" performing the methods described in the article is available at \url{https://github.com/nchenderson/ranefranking}. 
\end{description}
\vspace{-.6cm}

\bibliographystyle{agsm}
\bibliography{sdaarem}

\newpage 

\appendix
\setcounter{table}{0}
\renewcommand{\thetable}{S\arabic{table}}

\section{Deriving  of Posterior Expected Population Percentile Formula}
Under the working model 
\begin{equation}
\begin{cases} 
Y_{k} = \mathbf{x}_{k}^{T}\bbeta + \varv_{k} + e_{k},\qquad 
k = 1, \ldots, K,  \nonumber \\
\varv_{k} \sim N(0, \tau^{2}), \quad e_{k} \sim N(0, \sigma_{k}^{2}), \quad \varv_{k} \indep e_{k} \nonumber 
\end{cases}
\end{equation}
described in Section 2 of the main manuscript, the conditional distribution of $v_{k}$ given $Y_{k}$ is
\begin{equation}
\varv_{k}|Y_{k} \sim \textrm{Normal}\Big( B_{k,\tau}(Y_{k} - \mathbf{x}_{k}^{T}\bbeta), B_{k,\tau}\sigma_{k}^{2} \Big), \nonumber
\end{equation}
where $B_{k,\tau} = \tau^{2}/(\sigma_{k}^{2} + \tau^{2})$. The posterior expectation of the population percentile $\rho_{k}$ is then given by
\begin{eqnarray}
R_{k}(\bbeta, \tau, Y_{k} ) &=& E_{W}\big\{ \rho_{k} \mid Y_{k} \big\} \nonumber \\
&=& E_{W}\Big\{ \Phi(\varv_{k}/\tau) \Big| Y_{k} \Big\} \nonumber \\
&=& \frac{1}{\lambda_{k}}\int_{-\infty}^{\infty} \Phi(x/\tau) \phi\Big( \frac{x - \alpha_{k}(\bbeta)}{\lambda_{k}} \Big) dx, \label{eq:pep_basic_integral}
\end{eqnarray}
where $\alpha_{k}(\bbeta) = B_{k, \tau}(Y_{k} - \mathbf{x}_{k}^{T}\bbeta)$ and
$\lambda_{k} = \sigma_{k}\sqrt{B_{k, \tau}}$. Using
$\Phi(x/t) = \tfrac{1}{\tau}\int_{-\infty}^{x} \phi(t/\tau)$, we can rewrite (\ref{eq:pep_basic_integral}) as
\begin{eqnarray}
R_{k}(\bbeta, \tau, Y_{k})
&=& \frac{1}{\tau\lambda_{k}}\int_{-\infty}^{\infty} \int_{-\infty}^{x} \phi(t/\tau) \phi\Big( \frac{x - \alpha_{k}(\bbeta)}{\lambda_{k}} \Big) dtdx. \label{eq:pep_two_integral} 
\end{eqnarray}
To simplify (\ref{eq:pep_two_integral}), we consider the following change of variables
\begin{eqnarray}
w &=& x - \alpha_{k}(\bbeta) \nonumber \\
z &=& t - w. \nonumber 
\end{eqnarray}
This transformation has Jacobian matrix
\begin{equation}
J = \begin{bmatrix} 1 & 0 \\ 1 & 1 \end{bmatrix}, \nonumber
\end{equation}
and the region of integration $x \in \mathbb{R}, t \leq x$ in (\ref{eq:pep_two_integral}) will be transformed into the region of integration $w \in \mathbb{R}, z \leq \alpha_{k}(\bbeta)$
when using the transformed variables $(z, w)$. 
Hence, we can express the integral (\ref{eq:pep_two_integral}) as 
\begin{eqnarray}
R_{k}(\bbeta, \tau, Y_{k})
&=& \frac{1}{\tau\lambda_{k}}\int_{-\infty}^{\infty} \int_{-\infty}^{\alpha_{k}(\boldsymbol{\beta})} \textrm{det}(J) \phi\big((z + w)/\tau \big) \phi\big( w/\lambda_{k} \big) dzdw, \nonumber \\
&=& \frac{1}{\tau\lambda_{k}}\int_{-\infty}^{\alpha_{k}(\boldsymbol{\beta})} \int_{-\infty}^{\infty} \phi\big( (z + w)/\tau \big) \phi\big( w/\lambda_{k} \big) dwdz, 
\label{eq:order_change}
\end{eqnarray}
where the second equality follows from interchanging the order of integration and the fact that $\textrm{det}(J) = 1$.
To further simplify the integrand in (\ref{eq:order_change}), we can note that
\begin{eqnarray}
\phi((z + w)/\tau) \phi\big( w/\lambda_{k} \big)
&=& \frac{1}{2\pi}\exp\Big\{ - \frac{z^{2}}{2\tau^{2}} + \frac{2zw}{2\tau^{2}} - \frac{w^{2}}{2\tau^{2}} - \frac{w^{2}}{2\lambda_{k}^{2}} \Big\} \nonumber\\
&=& \frac{1}{\sqrt{2\pi}} \exp\Big\{ - \frac{z^{2}}{2\tau^{2}} \Big\}\frac{1}{2\pi}\exp\Big\{ \frac{2zw}{2\tau^{2}} - \frac{w^{2}(\lambda_{k}^{2} + \tau^{2})}{2\tau^{2}\lambda_{k}^{2}}  \Big\} \nonumber \\
&=& \frac{1}{\sqrt{2\pi}}\exp\Big\{- \frac{(\lambda_{k}^{2} + \tau^{2})}{2\tau^{2}\lambda_{k}^{2}}\Big( w^{2} - \frac{2zw\lambda_{k}^{2}}{ \lambda_{k}^{2} + \tau^{2}} + \frac{z^{2}\lambda_{k}^{4}}{ (\lambda_{k}^{2} + \tau^{2})^{2}} \Big) \Big\} \nonumber \\
&& \times \frac{1}{\sqrt{2\pi}} \exp\Big\{ - \frac{z^{2}}{2\tau^{2}} \Big\} \exp\Big\{ \frac{z^{2}\lambda_{k}^{2}}{ 2\tau^{2}(\lambda_{k}^{2} + \tau^{2})} \Big\} \nonumber \\
&=& \frac{1}{\sqrt{2\pi}}\exp\Big\{ -\frac{z^{2}(\lambda_{k}^{2} + \tau^{2})}{2\tau^{2}(\lambda_{k}^{2} + \tau^{2})} + \frac{z^{2}\lambda_{k}^{2}}{ 2\tau^{2}(\lambda_{k}^{2} + \tau^{2})} \Big\}\phi\Big( (w - z\theta_{k})/\kappa_{k} \Big) \nonumber \\
&=& \frac{1}{\sqrt{2\pi}} \exp\Big\{ -\frac{z^{2}}{ 2(\lambda_{k}^{2} + \tau^{2})} \Big\} \phi\Big( (w - \theta_{k}z)/\kappa_{k} \Big) \nonumber \\
&=& \phi\Big( z/\sqrt{\lambda_{k}^{2} + \tau^{2}} \Big)\phi\Big( (w - \theta_{k}z)/\kappa_{k} \Big), \label{eq:simplified_integrand}
\end{eqnarray}
where $\theta_{k} = \lambda_{k}^{2}/(\lambda_{k}^{2} + \tau^{2})$ and $\kappa_{k} = \tau\lambda_{k}/(\lambda_{k}^{2} + \tau^{2})^{1/2}$.

\noindent
Now, if we plug (\ref{eq:simplified_integrand}) into the integral (\ref{eq:order_change}), we have the following form of $R_{k}(\bbeta, \tau, Y_{k})$
\begin{eqnarray}
R_{k}(\bbeta, \tau, Y_{k})
&=& \frac{\kappa_{k}\sqrt{\lambda_{k}^{2} + \tau^{2}}}{\tau\lambda_{k}}\int_{-\infty}^{\alpha_{k}(\boldsymbol{\beta})} \frac{1}{\kappa_{k}\sqrt{\lambda_{k}^{2} + \tau^{2}}}\int_{-\infty}^{\infty}  \phi\Big( z/\sqrt{\lambda_{k}^{2} + \tau^{2}} \Big)\phi\Big( (w - \theta_{k}z)/\kappa_{k} \Big) dw dz\nonumber \\
&=&  \frac{\kappa_{k}\sqrt{\lambda_{k}^{2} + \tau^{2}}}{\tau\lambda_{k}}\int_{-\infty}^{\alpha_{k}(\boldsymbol{\beta})} \frac{1}{\sqrt{\lambda_{k}^{2} + \tau^{2}}}\phi\Big( \frac{z}{ \sqrt{\lambda_{k}^{2} + \tau^{2}} } \Big) dz \nonumber \\
&=& \frac{\kappa_{k}\sqrt{ \lambda_{k}^{2} + \tau^{2} }}{\tau\lambda_{k}}\Phi\Big( \frac{\alpha_{k}(\bbeta)}{ \sqrt{\lambda_{k}^{2} + \tau^{2}} } \Big) \nonumber \\
&=& \frac{\kappa_{k}\sqrt{ \lambda_{k}^{2} + \tau^{2} }}{\tau\lambda_{k}}\Phi\Big( \frac{B_{k,\tau}}{\sqrt{B_{k,\tau}(2\sigma_{k}^{2} + \tau^{2})}}(Y_{k} - \mathbf{x}_{k}^{T}\bbeta) \Big),
\label{eq:seclast_integral}
\end{eqnarray}
where the last equality uses the fact that $\alpha_{k}(\bbeta) = B_{k,\tau}(Y_{k} - \mathbf{x}_{k}^{T}\bbeta)$ and $\lambda_{k}^{2} + \tau^{2} = \sigma_{k}^{2}B_{k,\tau} + (\sigma_{k}^{2} + \tau^{2})B_{k,\tau} = B_{k,\tau}(2\sigma_{k}^{2} + \tau^{2})$.
Then, because $\kappa_{k} = \tau\lambda_{k}/(\lambda_{k}^{2} + \tau^{2})^{1/2}$, we can note that
\begin{equation}
\frac{\kappa_{k}\sqrt{ \lambda_{k}^{2} + \tau^{2} }}{\tau\lambda_{k}}
= \Big( \frac{\tau\lambda_{k}}{ \sqrt{\lambda_{k}^{2} + \tau^{2}} } \Big)\Big(\frac{\sqrt{ \lambda_{k}^{2} + \tau^{2} }}{\tau\lambda_{k}} \Big) = 1. 
\label{eq:kappa_equality}
\end{equation}
Combining (\ref{eq:seclast_integral}) and (\ref{eq:kappa_equality}), we can rewrite $R_{k}(\bbeta, \tau, Y_{k})$ as
\begin{equation}
R_{k}(\bbeta, \tau, Y_{k}) = \Phi\Big( \sqrt{\frac{B_{k,\tau}}{(2\sigma_{k}^{2} + \tau^{2})}}(Y_{k} - \mathbf{x}_{k}^{T}\bbeta) \Big) = \Phi\Big( V_{k,\tau}(Y_{k} - \mathbf{x}_{k}^{T}\bbeta) \Big), \nonumber
\end{equation}
where $V_{k,\tau} = \sqrt{B_{k,\tau}/(2\sigma_{k}^{2} + \tau^{2})}$.

\noindent
\textbf{Alternative Derivation.} As an alternative to deriving the form of $R_{k}(\bbeta, \tau, Y_{k})$,
we can note from (\ref{eq:pep_basic_integral}) that 
\begin{eqnarray}
R_{k}(\bbeta, \tau, Y_{k} )
= \frac{1}{\lambda_{k}}\int_{-\infty}^{\infty} \Phi(x/\tau) \phi\Big( \frac{x - \alpha_{k}(\bbeta)}{\lambda_{k}} \Big) dx 
= P(Z \leq X), \nonumber 
\end{eqnarray}
where $Z \sim \textrm{Normal}(0, \tau^{2})$, $Y \sim \textrm{Normal}(\alpha_{k}(\bbeta), \lambda_{k}^{2})$, 
and $Y$ and $Z$ are independent. Hence $Z - X \sim \textrm{Normal}( -\alpha_{k}(\bbeta), \tau^{2} + \lambda_{k}^{2})$,
and we can express $P(Z \leq X)$ as
\begin{eqnarray}
R_{k}(\bbeta, \tau, Y_{k} ) &=& P(Z \leq X) \nonumber \\
&=& P(Z - X \leq 0) \nonumber \\
&=& P\Big( \frac{Z - X + \alpha_{k}(\bbeta)}{\sqrt{ \tau^{2} + \lambda_{k}^{2} } } \leq \frac{\alpha_{k}(\bbeta)}{\sqrt{ \tau^{2} + \lambda_{k}^{2} } } \Big) \nonumber \\
&=& \Phi\Big(\frac{\alpha_{k}(\bbeta)}{\sqrt{ \tau^{2} + \lambda_{k}^{2} } } \Big). \label{eq:Ralternative}
\end{eqnarray}
Because $\alpha_{k}(\bbeta) = B_{k,\tau}(Y_{k} - \mathbf{x}_{k}^{T}\bbeta)$ and $\lambda_{k}^{2} + \tau^{2} = B_{k,\tau}(2\sigma_{k}^{2} + \tau^{2})$, we can re-write (\ref{eq:Ralternative}) as
\begin{equation}
R_{k}(\bbeta, \tau, Y_{k} ) = \Phi\Big(\frac{B_{k,\tau}(Y_{k} - \mathbf{x}_{k}^{T}\bbeta)}{\sqrt{ B_{k,\tau}(2\sigma_{k}^{2} + \tau^{2}) } } \Big) = \Phi\Big( V_{k,\tau}(Y_{k} - \mathbf{x}_{k}^{T}\bbeta)  \Big). \nonumber 
\end{equation}

\section{Proof of Theorem 1: Unbiasedness of $\hat{Q}_{\tau}(\boldsymbol{\beta})$}

\begin{lemma}
\label{lem:ubiased}
If $(Y_{k}, \varv_{k})$ have the joint distribution described in the main manuscript, then
the following holds for any integer $h \geq 0$ and infinitely differentiable function $g:\mathbb{R} \longrightarrow \mathbb{R}$ 
\begin{equation}
E\Big\{ \varv_{k}^{h+1} g(Y_{k}) \Big\}
= \sum_{j=0}^{\lfloor (h+1)/2 \rfloor } \frac{(h + 1)!}{2^{j}(h + 1 - 2j)!j!}(\tau^{2})^{h + 1 - j}E\Big\{ \frac{\partial^{h - 2j + 1} g(Y_{k})}{\partial Y_{k} } \Big\}. \label{eq:ubform_lem1}
\end{equation}
\end{lemma}

\bigskip

\noindent
\textit{Proof:} 
We will prove that (\ref{eq:ubform_lem1}) holds for any $h \geq 0$ through induction.
For the case of $h = 0$, the formula (\ref{eq:ubform_lem1}) states that
\begin{equation}
E\Big\{ \varv_{i}^{h+1} g(Y_{i}) \Big\} = E\Big\{ \varv_{i} g(Y_{i}) \Big\}
= \sum_{j=0}^{0} \frac{(h + 1)!}{2^{j}(h + 1 - 2j)!j!}(\tau^{2})^{1 - j}E\Big\{ \frac{\partial^{1 - 2j} g(Y_{i})}{\partial Y_{i} } \Big\}
= \tau^{2}E\Big\{ \frac{\partial g(Y_{i})}{\partial Y_{i} } \Big\}. \nonumber 
\end{equation}
which is just a re-statement of a multivariate version of Stein's lemma (e.g., Lemma 1 in \cite{liu1994}).

\medskip

\noindent
\textbf{Inductive step for odd h.} Assume that (\ref{eq:ubform_lem1}) is true for an integer $h > 0$ and that $h$ is odd. The inductive step for even $h$ is nearly identical and we leave it omitted.

\medskip

\noindent
To further examine $E\big\{ v_{i}^{h+1} g(Y_{i}) \big\}$, first note that it follows from a multivariate version of Stein's lemma (e.g., Lemma 1 in \cite{liu1994}) that
\begin{equation}
E\Big\{ \varv_{i}^{h + 2} g(Y_{i}) \Big\}
= E\Big\{ \varv_{i}\varv_{i}^{h + 1} g(Y_{i}) \Big\}
= \tau^{2}(h+1)E\Big\{ \varv_{i}^{h} g(Y_{i}) \Big\}
+ \tau^{2}E\Big\{ \varv_{i}^{h+1} \frac{\partial g(Y_{i})}{\partial Y_{i}} \Big\}.
\label{eq:isserlis_recursion}
\end{equation}
Note that the multivariate Stein's lemma applied to our setting implies that for a function $f: \mathbb{R}^{2} \longrightarrow \mathbb{R}$, we have $E\{ \varv_{i} f(\varv_{i}, Y_{i}) \} = \textrm{Cov}(\varv_{i}, \varv_{i})E\{ \partial f(\varv_{i}, Y_{i})/\partial \varv_{i}\} + \textrm{Cov}(\varv_{i}, Y_{i})E\{ \partial f(\varv_{i}, Y_{i})/\partial \varv_{i}\}$.

\medskip

\noindent
Now, assuming that (\ref{eq:ubform_lem1}) is true for $h$, we can re-write (\ref{eq:isserlis_recursion}) as
\begin{eqnarray}
E\Big\{ \varv_{i}^{h + 2} g(Y_{i}) \Big\}
&=& \tau^{2}(h+1)\sum_{j=0}^{\lfloor h/2 \rfloor} \frac{h!}{(h - 2j)!(2j)!}(\tau^{2})^{h - j}E\Big\{ \frac{\partial^{h - 2j} g(Y_{i})}{\partial Y_{i} } \Big\} \nonumber \\
&+& \tau^{2}\sum_{j=0}^{\lfloor (h+1)/2 \rfloor } \frac{(h + 1)!}{(h - 2j + 1)!(2j)!}(\tau^{2})^{h - j + 1}E\Big\{ \frac{\partial^{h - 2j + 2} g(Y_{i})}{\partial Y_{i} } \Big\}  \nonumber \\
&=& (\tau^{2})^{h + 2}E\Big\{ \frac{\partial^{h + 2} g(Y_{i})}{\partial Y_{i} } \Big\} + \sum_{j=0}^{ (h-1)/2} \frac{(h+1)!}{2^{j}(h - 2j)!j!}(\tau^{2})^{h - j + 1}E\Big\{ \frac{\partial^{h - 2j} g(Y_{i})}{\partial Y_{i} } \Big\} \nonumber \\
&+& \sum_{j=1}^{(h+1)/2 } \frac{(h + 1)!}{2^{j}(h - 2j + 1)!j!} (\tau^{2})^{h - j + 2} E\Big\{ \frac{\partial^{h - 2j + 2} g(Y_{i})}{\partial Y_{i} } \Big\}  \nonumber \\
&=& \sum_{j=0}^{ (h-1)/2} \frac{2(j+1)(h+2)!(\tau^{2})^{h + 2 - (j + 1)}}{(h+2)2^{j+1}(h + 2 - 2(j+1))!(j+1)!}E\Big\{ \frac{\partial^{h + 2 - 2(j + 1)} g(Y_{i})}{\partial Y_{i} } \Big\} \nonumber \\
&+& \sum_{j=1}^{(h+1)/2 } \frac{(h + 1)!}{2^{j}(h - 2j + 1)!j!}(\tau^{2})^{h - j + 2}E\Big\{ \frac{\partial^{h - 2j + 2} g(Y_{i})}{\partial Y_{i} } \Big\} + (\tau^{2})^{h + 2}E\Big\{ \frac{\partial^{h + 2} g(Y_{i})}{\partial Y_{i} } \Big\} \nonumber \\
&=& \sum_{u=1}^{ (h+1)/2 } \frac{2u(h+2)!(\tau^{2})^{h - u + 2}}{(h+2)2^{u}(h + 2 - 2u)!u!}E\Big\{ \frac{\partial^{h - 2u + 2} g(Y_{i})}{\partial Y_{i} } \Big\} \nonumber \\
&+& \sum_{j=1}^{(h+1)/2 } \frac{(h + 1)!}{2^{j}(h - 2j + 1)!j!}(\tau^{2})^{h - j + 2}E\Big\{ \frac{\partial^{h - 2j + 2} g(Y_{i})}{\partial Y_{i} } \Big\} + (\tau^{2})^{h + 2}E\Big\{ \frac{\partial^{h + 2} g(Y_{i})}{\partial Y_{i} } \Big\} \nonumber \\
&=& \sum_{j=1}^{(h+1)/2 } (\tau^{2})^{h - j + 2}\Bigg(\frac{2j(h+2)!}{(h+2)2^{j}(h + 2 - 2j)!j!} + \frac{(h + 1)!}{2^{j}(h - 2j + 1)!j!} \Bigg)E\Big\{ \frac{\partial^{h - 2j + 2} g(Y_{i})}{\partial Y_{i} } \Big\} \nonumber \\
&+& (\tau^{2})^{h + 2}E\Big\{ \frac{\partial^{h + 2} g(Y_{i})}{\partial Y_{i} } \Big\}. 
\label{eq:main_induct_even}
\end{eqnarray}
Now, note that
\begin{eqnarray}
\Bigg(\frac{2j(h+2)!}{(h+2)2^{j}(h + 2 - 2j)!j!} + \frac{(h + 1)!}{2^{j}(h - 2j + 1)!j!} \Bigg)
&=& \frac{2j(h+2)! + (h + 2)!(h + 2 - 2j) }{(h+2)2^{j}(h + 2 - 2j)!j!} \nonumber \\
&=& \frac{(h+2)![2j + (h + 2 - 2j)] }{(h+2)2^{j}(h + 2 - 2j)!j!} \nonumber \\
&=& \frac{(h+2)!}{2^{j}(h + 2 - 2j)!j!}.
\label{eq:key_simplification}
\end{eqnarray}
Then, if we plug (\ref{eq:key_simplification}) back into (\ref{eq:main_induct_even}), we obtain
\begin{eqnarray}
E\Big\{ \varv_{i}^{h + 2} g(Y_{i}) \Big\}
&=& \sum_{j=1}^{(h+1)/2 } (\tau^{2})^{h - j + 2}\Bigg( \frac{(h+2)!}{2^{j}(h + 2 - 2j)!j!} \Bigg)E\Big\{ \frac{\partial^{h - 2j + 2} g(Y_{i})}{\partial Y_{i} } \Big\} + (\tau^{2})^{h + 2}E\Big\{ \frac{\partial^{h + 2} g(Y_{i})}{\partial Y_{i} } \Big\}
\nonumber \\
&=& \sum_{j=0}^{(h+1)/2 } (\tau^{2})^{h - j + 2}\Bigg( \frac{(h+2)!}{2^{j}(h + 2 - 2j)!j!} \Bigg)E\Big\{ \frac{\partial^{h - 2j + 2} g(Y_{i})}{\partial Y_{i} } \Big\} \nonumber \\
&=& \sum_{j=0}^{ \lfloor (h+2)/2 \rfloor } (\tau^{2})^{h - j + 2}\Bigg( \frac{(h+2)!}{2^{j}(h + 2 - 2j)!j!} \Bigg)E\Big\{ \frac{\partial^{h - 2j + 2} g(Y_{i})}{\partial Y_{i} } \Big\}, \nonumber 
\end{eqnarray}
which verifies the desired inequality.

\subsection{Proof of Theorem 1} 
To prove Theorem 1 in the main manuscript, we can first note that
\begin{equation}
\rho_{k} - 1/2 = \Phi( \varv_{k}/\tau) - 1/2 
= \frac{1}{\sqrt{2\pi}}\sum_{h=0}^{\infty} \frac{(-1)^{h}(\tau)^{-(2h + 1)}\varv_{k}^{(2h + 1)}}{2^{h}h!(2h + 1)}, \nonumber
\end{equation}
which implies that $Q_{\tau}(\bbeta)$ can be expressed as
\begin{eqnarray}
Q_{\tau}(\bbeta) 
&=& \frac{1}{12} - \frac{2}{K}\sum_{k=1}^{K}E\Big\{ (\rho_{k} - 1/2)R_{k}(\bbeta, \tau, Y_{k}) \Big\}
+ \frac{1}{K}\sum_{k=1}^{K} E\Big\{ \Big( R_{k}(\bbeta, \tau, Y_{k}) - \frac{1}{2}\Big)^{2} \Big\} \nonumber \\
&=& \frac{1}{12} - \frac{2}{K}\sum_{k=1}^{K} \sum_{h=0}^{\infty} \frac{(-1)^{h}(\tau^{2})^{-(2h + 1)}}{2^{h}h!(2h + 1)} E\Big\{ \varv_{k}^{(2h + 1)}R_{k}(\bbeta, \tau, Y_{k}) \Big\} \nonumber \\
&& + \frac{1}{K}\sum_{k=1}^{K} E\Big\{ \Big( R_{k}(\bbeta, \tau, Y_{k}) - \frac{1}{2}\Big)^{2} \Big\}. \nonumber
\end{eqnarray}
Applying Lemma \ref{lem:ubiased} to the above gives
\begin{eqnarray}
Q_{\tau}(\bbeta) 
&=& \frac{1}{12} - \frac{2}{K}\sum_{k=1}^{K} \sum_{h=0}^{\infty} \frac{(-1)^{h}(\tau^{2})^{-(h + 1/2)}}{2^{h}h!(2h + 1)} \sum_{j=0}^{h } \frac{(2h + 1)!}{2^{j}(2h + 1 - 2j)!j!}(\tau^{2})^{2h + 1 - j}E\Big\{ \frac{\partial^{2h - 2j + 1} R_{k}(\bbeta, \tau, Y_{k})}{\partial Y_{k} } \Big\} \nonumber \\
&& + \frac{1}{K}\sum_{k=1}^{K} E\Big\{ \Big( R_{k}(\bbeta, \tau, Y_{k}) - \frac{1}{2}\Big)^{2} \Big\}.
\label{eq:Qexp}
\end{eqnarray}
From (\ref{eq:Qexp}), it is clear that the following is an unbiased estimator
of $Q_{\tau}(\bbeta)$:
\begin{eqnarray}
\hat{Q}_{\tau}(\bbeta) 
&=& \frac{1}{12} - \frac{2}{K}\sum_{k=1}^{K} \sum_{h=0}^{\infty} \frac{(-1)^{h}(\tau^{2})^{-(h + 1/2)}}{2^{h}h!(2h + 1)} \sum_{j=0}^{h } \frac{(2h + 1)!}{2^{j}(2h + 1 - 2j)!j!}(\tau^{2})^{2h + 1 - j}\frac{\partial^{2h - 2j + 1} R_{k}(\bbeta, \tau, Y_{k})}{\partial Y_{k} }  \nonumber \\
&& + \frac{1}{K}\sum_{k=1}^{K} \Big( R_{k}(\bbeta, \tau, Y_{k}) - \frac{1}{2}\Big)^{2}. \nonumber
\end{eqnarray}

\section{Proof of Proposition 1}
To demonstrate that Proposition 1 holds, we will use the following rearrangement inequality which can be found, for example, in \citep{hardy1988}.
\begin{lemma}
Rearrangement Inequality \citep{hardy1988}: Consider $a_{k} \in \mathbb{R}$, $b_{k} \in \mathbb{R}$ satisfying $a_{1} \leq a_{2} \leq \ldots \leq a_{K}$ and $b_{1} \leq b_{2} \leq \ldots \leq b_{K}$. Then, the following inequality holds for any permutation $b_{(1)}, b_{(2)}, \ldots, b_{(K)}$ of $b_{1}, b_{2}, \ldots, b_{K}$:
\begin{equation}
\sum_{k=1}^{K} a_{k}b_{(k)} \leq \sum_{k=1}^{K} a_{k}b_{k}. \nonumber
\end{equation}
\end{lemma}

\medskip

\noindent
In Proposition 1, our aim was to minimize, $\sum_{k=1}^{K} \big\{ \tfrac{1}{K+1}A_{k} - R_{k}(\hatbbeta_{r}, \tau, Y_{k}) \big\}^{2}$, subject to the constraint that $A_{1}, \ldots, A_{K}$ exhaust the integers $1, \ldots, K$. Letting 
$\tilde{R}_{k}(\hatbbeta_{r}, \tau, Y_{k}) = (K+1) R_{k}(\hatbbeta_{r}, \tau, Y_{k})$, this aim is equivalent to minimizing
\begin{equation}
f(A_{1}, \ldots, A_{K}) = \sum_{k=1}^{K} \{ A_{k} - \tilde{R}_{k}(\hatbbeta_{r}, \tau, Y_{k}) \}^{2}, 
\label{eq:perm_obj}
\end{equation}
where $A_{1}, \ldots, A_{K}$ is a permutation of the set of integers $\{ 1, 2, \ldots, K\}$. 
We can assume without loss of generality that
$\tilde{R}_{1}(\hatbbeta_{r}, \tau, Y_{1}) \leq \tilde{R}_{2}(\hatbbeta_{r}, \tau, Y_{2})
\leq \ldots \leq \tilde{R}_{K}(\hatbbeta_{r}, \tau, Y_{K})$. Then, 
because $A_{1}, \ldots, A_{K}$ is assumed to be a permutation of $\{1, \ldots, K\}$, we can rewrite
(\ref{eq:perm_obj}) as
\begin{eqnarray}
f(A_{1}, \ldots, A_{K}) &=& \sum_{k=1}^{K} A_{k}^{2} - 2\sum_{k=1}^{K} A_{k}\tilde{R}_{k}(\hatbbeta_{r}, \tau, Y_{k}) 
+ \sum_{k=1}^{K} \tilde{R}_{k}^{2}(\hatbbeta_{r}, \tau, Y_{k}) \nonumber \\
&=& \frac{2K^{3} + 3K^{2} + K}{6} + \sum_{k=1}^{K} \tilde{R}_{k}^{2}(\hatbbeta_{r}, \tau, Y_{k})
- 2\sum_{k=1}^{K} A_{k}\tilde{R}_{k}(\hatbbeta_{r}, \tau, Y_{k}).
\label{eq:perm_obj_simplify}
\end{eqnarray}
If $A_{k}^{*}$ is a permutation of $\{1, \ldots, K\}$ satisfying $A_{1}^{*} \leq A_{2}^{*} \leq \ldots \leq A_{K}^{*}$,
then it follows from (\ref{eq:perm_obj_simplify}) and the rearrangement inequality that
\begin{equation}
f(A_{1}, \ldots, A_{K}) 
\leq \frac{2K^{3} + 3K^{2} + K}{6} + \sum_{k=1}^{K} \tilde{R}_{k}^{2}(\hatbbeta_{r}, \tau, Y_{k})
- 2\sum_{k=1}^{K} A_{k}^{*}\tilde{R}_{k}(\hatbbeta_{r}, \tau, Y_{k}). \nonumber
\end{equation}
That is, the $A_{k}^{*}$ must have the same ordering as the $\tilde{R}_{k}(\hatbbeta_{r}, \tau, Y_{k})$.

\section{Proof of Theorem 2: Monotonicity of the MM Algorithm}

Before proving Theorem 2 in the main manuscript, we first prove the following two lemmas.
\begin{lemma}
\label{lem:mm1}
Consider the functions $\tilde{g}_{k}: \mathbb{R}^{p} \longrightarrow \mathbb{R}$ defined as
\begin{equation}
\tilde{g}_{k}(\bbeta) = 1 - \Big\{ \Phi\Big( V_{k,\tau}(Y_{k} - \mathbf{x}_{k}^{T}\bbeta) \Big) - 1/2\Big\}^{2}, \quad k = 1, \ldots, K. \nonumber
\end{equation}
Then for any $\bbeta^{(t)} \in \mathbb{R}^{p}$ and nonnegative weights $w_{1}, \ldots, w_{K}$, we have
\begin{eqnarray}
-\sum_{k=1}^{K} w_{k} \log(\tilde{g}_{k}(\bbeta)) &\leq& \tilde{G}(\bbeta|\bbeta^{(t)}, \mathbf{w}) \nonumber \\
-\sum_{k=1}^{K} w_{k} \log(\tilde{g}_{k}(\bbeta^{(t)})) &=& \tilde{G}(\bbeta^{(t)}|\bbeta^{(t)}, \mathbf{w}), \nonumber 
\end{eqnarray}
where $\tilde{G}(\bbeta^{(t)}|\bbeta^{(t)}, \mathbf{w})$ is defined as
\begin{eqnarray}
\tilde{G}(\bbeta | \bbeta^{(t)}, \mathbf{w})
&=& -\sum_{k=1}^{K} w_{k} \log(\tilde{g}_{k}(\bbeta^{(t)})) + \nabla \varphi(\bbeta^{(t)})^{T}(\bbeta - \bbeta^{(t)})
+ \frac{1}{2}(\bbeta - \bbeta^{(t)})^{T}\tfrac{1}{6}\mathbf{X}^{T}\mathbf{V}^{T}\mathbf{w}\mathbf{V}\mathbf{X}(\bbeta - \bbeta^{(t)}) \nonumber \\
&=& C_{\tilde{G}}^{(t)} + \Big[\nabla \varphi(\bbeta^{(t)})^{T}
- \frac{1}{6}(\bbeta^{(t)})^{T}\mathbf{X}^{T}\mathbf{V}^{T}\mathbf{w}\mathbf{V}\mathbf{X} \Big]\bbeta
+ \frac{1}{12}\bbeta^{T}\mathbf{X}^{T}\mathbf{V}^{T}\mathbf{w}\mathbf{V}\mathbf{X}\bbeta \nonumber \\
&=& C_{\tilde{G}}^{(t)}(\mathbf{w}) - \bbeta^{T}\Big[\mathbf{X}^{T}\mathbf{V}^{T}\mathbf{W}\mathbf{d}(\bbeta^{(t)})
+ \frac{1}{6}\mathbf{X}^{T}\mathbf{V}^{T}\mathbf{W}\mathbf{V}\mathbf{X}\bbeta^{(t)} \Big]
+ \frac{1}{12}\bbeta^{T}\mathbf{X}^{T}\mathbf{V}^{T}\mathbf{W}\mathbf{V}\mathbf{X}\bbeta, \nonumber
\end{eqnarray}
where $\mathbf{W} = \textrm{diag}\{ w_{1}, \ldots, w_{K} \}$, $\mathbf{V} = \textrm{diag}\{ V_{1,\tau}, \ldots, V_{K,\tau} \}$, and where $\mathbf{d}(\bbeta^{(t)}) \in \mathbb{R}^{K}$ is the vector with components
\begin{equation}
d_{k}(\bbeta^{(t)}) = \frac{2\phi(V_{k,\tau}(Y_{k} - \mathbf{x}_{k}^{T}\bbeta^{(t)})D( V_{k,\tau}(Y_{k} - \mathbf{x}_{k}^{T}\bbeta^{(t)}) ) }{ 1 - D^{2}( V_{k,\tau}(Y_{k} - \mathbf{x}_{k}^{T}\bbeta^{(t)}) ) }, \nonumber
\end{equation}
where $D$ is the function defined as $D(u) = \Phi(u) - 1/2$, and the constant $C_{\tilde{G}}^{(t)}(\mathbf{w})$ is defined as
\begin{equation}
C_{\tilde{G}}^{(t)}(\mathbf{w}) = -\sum_{k=1}^{K} w_{k} \log\{ \tilde{g}_{k}(\bbeta^{(t)}) \} +  d(\bbeta^{(t)})^{T}\mathbf{V}^{T}\mathbf{W}\mathbf{X}\bbeta^{(t)}
+ \frac{1}{12}(\bbeta^{(t)})^{T}\mathbf{X}^{T}\mathbf{V}^{T}\mathbf{W}\mathbf{V}\mathbf{X}\bbeta^{(t)}. \nonumber 
\end{equation}
\end{lemma}

\medskip

\noindent
\textit{Proof.}
Define $\varphi(\bbeta)$ as the sum $\varphi(\bbeta) = -\sum_{k=1}^{K}  w_{k}\log(\tilde{g}_{k}(\bbeta))$, and first note that the components of the gradient of $\varphi(\bbeta)$ are given by
\begin{equation}
\frac{\partial \varphi(\bbeta)}{\partial \beta_{h}} = -\sum_{k=1}^{K} w_{k} V_{k,\tau}x_{kh}d\Big(V_{k,\tau}(Y_{k} - \mathbf{x}_{k}^{T}\bbeta) \Big), \nonumber
\end{equation}
where the function $d$ is defined as
\begin{equation}
d(u) = \frac{2\phi(u)\{\Phi(u) - 1/2\} }{1 - \{\Phi(u) -1/2\}^{2} }. \nonumber 
\end{equation}
The derivative of $d$ is given by
\begin{equation}
d'(u) = 2\phi(u)\Big(\frac{\phi(u)[1 + \{\Phi(u) - 1/2\}^{2}] }{[1 - \{\Phi(u) -1/2\}^{2}]^{2} } - \frac{u\{\Phi(u) - 1/2\} }{1 - \{\Phi(u) -1/2\}^{2} } \Big). \nonumber 
\end{equation}
The components of the Hessian of $\varphi(\bbeta)$ are given by
\begin{equation}
\frac{\partial^{2} \varphi(\bbeta)}{\partial \beta_{h}\beta_{j}} = \sum_{k=1}^{K} w_{k}V_{k,\tau}^{2}x_{kh}x_{kj}d'\Big(V_{k,\tau}(Y_{k} - \mathbf{x}_{k}^{T}\bbeta) \Big), \nonumber
\end{equation}
and hence, we can express the Hessian in matrix form as
\begin{equation}
\nabla^{2} \varphi(\bbeta) = \mathbf{X}^{T}\mathbf{V}^{T}\mathbf{W}\mathbf{D}_{1}(\bbeta)\mathbf{V}\mathbf{X}, \nonumber
\end{equation}
where $\mathbf{W} = \textrm{diag}\{w_{1}, \ldots, w_{K} \}$ and 
\begin{equation}
D_{1}(\bbeta) = \textrm{diag}\Big\{ d'\Big(V_{k,\tau}(Y_{k} - \mathbf{x}_{k}^{T}\bbeta) \Big), \ldots, d'\Big(V_{k,\tau}(Y_{k} - \mathbf{x}_{k}^{T}\bbeta) \Big) \Big\}. \nonumber
\end{equation}
One can show that $d'(u) \leq 1/3$ for any value of $u$ and hence,
\begin{eqnarray}
\tfrac{1}{3}\mathbf{X}^{T}\mathbf{V}^{T}\mathbf{W}\mathbf{V}\mathbf{X} - \nabla^{2} \varphi(\bbeta) 
&=& \tfrac{1}{3}\mathbf{X}^{T}\mathbf{V}^{T}\mathbf{W}\mathbf{V}\mathbf{X} - \mathbf{X}^{T}\mathbf{V}^{T}\mathbf{W}\mathbf{D}_{1}(\bbeta)\mathbf{V}\mathbf{X}  \nonumber \\
&=& \mathbf{X}^{T}\mathbf{V}^{T}(\tfrac{1}{3}\mathbf{W} - \mathbf{W}\mathbf{D}_{1}(\bbeta))\mathbf{V}\mathbf{X} \nonumber \\
&\geq& 0. \nonumber
\end{eqnarray}
If follows from the fact that $\tfrac{1}{3}\mathbf{X}^{T}\mathbf{V}^{T}\mathbf{W}\mathbf{V}\mathbf{X} - \nabla^{2} \varphi(\bbeta) \geq 0$, that for any $\bbeta$, we have (e.g., \cite{lange2010})
\begin{eqnarray}
\varphi(\bbeta) &\leq& \varphi(\bbeta^{(t)}) + \nabla \varphi(\bbeta^{(t)})^{T}(\bbeta - \bbeta^{(t)})
+ \frac{1}{2}(\bbeta - \bbeta^{(t)})^{T}\tfrac{1}{3}\mathbf{X}^{T}\mathbf{V}^{T}\mathbf{W}\mathbf{V}\mathbf{X}(\bbeta - \bbeta^{(t)}) \nonumber \\
&=& \tilde{G}(\bbeta|\bbeta^{(t)}). \nonumber
\end{eqnarray}
From the above definition of $\tilde{G}(\bbeta|\bbeta^{(t)})$, it is also clear
that $\tilde{G}(\bbeta^{(t)}|\bbeta^{(t)}) = \varphi(\bbeta^{(t)})$.

\bigskip

\begin{lemma}
\label{lem:mm2}
For nonnegative functions $g_{k}:\mathbb{R}^{p} \longrightarrow \mathbb{R}$ and 
$f_{k}:\mathbb{R}^{p} \longrightarrow \mathbb{R}$ and $h_{k} = f_{k} + g_{k}$, the following inequality holds
for any vectors $\bbeta \in \mathbb{R}^{p}$ and $\bbeta^{(t)} \in \mathbb{R}^{p}$:
\begin{eqnarray}
-\log\Big( \sum_{k=1}^{K} h_{k}(\bbeta) \Big) &\leq& -\sum_{k=1}^{K} W_{k1}(\bbeta^{(t)})\log\Big( f_{k}(\bbeta) \Big) - \sum_{k=1}^{K} W_{k2}(\bbeta^{(t)})\log\Big( g_{k}(\bbeta) \Big) \nonumber \\
&+& \sum_{k=1}^{K} W_{k1}(\bbeta^{(t)})\log\Big( w_{k1}(\bbeta^{(t)}) \Big) + \sum_{k=1}^{K} W_{k2}(\bbeta^{(t)})\log\Big( w_{k2}(\bbeta^{(t)}) \Big) \nonumber \\
&+& \sum_{k=1}^{K} W_{k}(\bbeta^{(t)})\log\Big( W_{k}(\bbeta^{(t)}) \Big), \nonumber 
\end{eqnarray}
where $W_{k}(\bbeta)$, $w_{k1}(\bbeta)$, and $w_{k2}(\bbeta)$ are defined as:
\begin{equation}
W_{k}(\bbeta) = \frac{ h_{k}(\bbeta) }{ \sum_{k=1}^{K} h_{k}(\bbeta) }, \qquad
w_{k1}(\bbeta) = \frac{ f_{k}(\bbeta) }{ h_{k}(\bbeta) }, \qquad 
w_{k2}(\bbeta) = \frac{ g_{k}(\bbeta) }{ h_{k}(\bbeta) }, \nonumber 
\end{equation}
and $W_{k1}(\bbeta)$ and $W_{k2}(\bbeta)$ are defined as
\begin{eqnarray}
W_{k1}(\bbeta) &=& W_{k}(\bbeta)w_{k1}(\bbeta) = \frac{f_{k}(\bbeta)}{ \sum_{k=1}^{K} h_{k}(\bbeta) } \nonumber \\
W_{k2}(\bbeta) &=& W_{k}(\bbeta)w_{k2}(\bbeta) = \frac{g_{k}(\bbeta)}{ \sum_{k=1}^{K} h_{k}(\bbeta) } \nonumber
\end{eqnarray}
\end{lemma}

\medskip

\noindent
\textit{Proof.}
Let $G(\bbeta) = -\log\{ \sum_{k=1}^{K} h_{k}(\bbeta) \}$ and first note that, because $\sum_{k=1}^{K} W_{k}(\bbeta^{(t)}) = 1$, we have
\begin{eqnarray}
G(\bbeta) &=& \sum_{k=1}^{K} W_{k}(\bbeta^{(t)})G(\bbeta) \nonumber \\
&=& -\sum_{k=1}^{K} W_{k}(\bbeta^{(t)})\log\Big( \frac{h_{k}(\bbeta)}{W_{k}(\bbeta)} \Big) \nonumber \\
&=& -\sum_{k=1}^{K} W_{k}(\bbeta^{(t)})\log\Big( h_{k}(\bbeta) \Big)
- \sum_{k=1}^{K} W_{k}(\bbeta^{(t)})\log\Big( \frac{1}{W_{k}(\bbeta)} \Big) \nonumber \\
&=& -\sum_{k=1}^{K} W_{k}(\bbeta^{(t)})\log\Big( h_{k}(\bbeta) \Big)
- \sum_{k=1}^{K} W_{k}(\bbeta^{(t)})\log\Big( \frac{W_{k}(\bbeta^{(t)})}{W_{k}(\bbeta)} \Big)
+ \sum_{k=1}^{K} W_{k}(\bbeta^{(t)})\log\Big( W_{k}(\bbeta^{(t)} \Big) \nonumber \\
&\leq&  -\sum_{k=1}^{K} W_{k}(\bbeta^{(t)})\log\Big( h_{k}(\bbeta) \Big)
+ \sum_{k=1}^{K} W_{k}(\bbeta^{(t)})\log\Big( W_{k}(\bbeta^{(t)} \Big),
\label{eq:lem2_firstinequality}
\end{eqnarray}
where the last inequality follows from Gibb's inequality.

\medskip

\noindent
Now, note that, because $w_{k1}(\bbeta^{(t)}) + w_{k2}(\bbeta^{(t)}) = 1$, we also have
\begin{eqnarray}
-\sum_{k=1}^{K} W_{k}(\bbeta^{(t)})\log\Big( h_{k}(\bbeta) \Big)
&=& -\sum_{k=1}^{K} W_{k}(\bbeta^{(t)})\Big[ w_{k1}(\bbeta^{(t)}) \log\Big( h_{k}(\bbeta) \Big)
+ w_{k2}(\bbeta^{(t)}) \log\Big( h_{k}(\bbeta) \Big) \Big] \nonumber \\
&=& -\sum_{k=1}^{K} W_{k}(\bbeta^{(t)})\Big[ w_{k1}(\bbeta^{(t)}) \log\Big( \frac{f_{k}(\bbeta)}{w_{k1}(\bbeta)} \Big)
+ w_{k2}(\bbeta^{(t)}) \log\Big( \frac{g_{k}(\bbeta)}{w_{k2}(\bbeta)} \Big) \Big] \nonumber \\
&=& \sum_{k=1}^{K} W_{k}(\bbeta^{(t)})\Big[ -w_{k1}(\bbeta^{(t)}) \log\Big( f_{k}(\bbeta) \Big)
- w_{k2}(\bbeta^{(t)}) \log\Big( g_{k}(\bbeta) \Big)  \nonumber \\
&& -w_{k1}(\bbeta^{(t)}) \log\Big( \frac{w_{k1}(\bbeta^{(t)})}{w_{k1}(\bbeta)} \Big)
- w_{k2}(\bbeta^{(t)}) \log\Big( \frac{w_{k2}(\bbeta^{(t)})}{w_{k2}(\bbeta)} \Big) \nonumber \\
&& + w_{k1}(\bbeta^{(t)}) \log\Big( w_{k1}(\bbeta^{(t)}) \Big)
+ w_{k2}(\bbeta^{(t)}) \log\Big( w_{k2}(\bbeta^{(t)}) \Big) \Big] \nonumber \\
&\leq& \sum_{k=1}^{K} W_{k}(\bbeta^{(t)})\Big[ -w_{k1}(\bbeta^{(t)}) \log\Big( f_{k}(\bbeta) \Big)
- w_{k2}(\bbeta^{(t)}) \log\Big( g_{k}(\bbeta) \Big)  \nonumber \\
&& + w_{k1}(\bbeta^{(t)}) \log\Big( w_{k1}(\bbeta^{(t)}) \Big)
+ w_{k2}(\bbeta^{(t)}) \log\Big( w_{k2}(\bbeta^{(t)}) \Big) \Big],
\label{eq:lem2_secondinequality}
\end{eqnarray}
where the last inequality again follows from Gibb's inequality.

\medskip

\noindent
By combining (\ref{eq:lem2_firstinequality}) and (\ref{eq:lem2_secondinequality}), we now have
\begin{eqnarray}
-\log\Big( \sum_{k=1}^{K} h_{k}(\bbeta) \Big)
&\leq&  \sum_{k=1}^{K} W_{k}(\bbeta^{(t)})\Big[ -w_{k1}(\bbeta^{(t)}) \log\Big( f_{k}(\bbeta) \Big)
- w_{k2}(\bbeta^{(t)}) \log\Big( g_{k}(\bbeta) \Big)  \nonumber \\
&& + w_{k1}(\bbeta^{(t)}) \log\Big( w_{k1}(\bbeta^{(t)}) \Big)
+ w_{k2}(\bbeta^{(t)}) \log\Big( w_{k2}(\bbeta^{(t)}) \Big) \Big] \nonumber \\
&& + \sum_{k=1}^{K} W_{k}(\bbeta^{(t)})\log\Big( W_{k}(\bbeta^{(t)}) \Big) \nonumber \\
&=& -\sum_{k=1}^{K} W_{k1}(\bbeta^{(t)})\log\Big( f_{k}(\bbeta) \Big) - \sum_{k=1}^{K} W_{k2}(\bbeta^{(t)})\log\Big( g_{k}(\bbeta) \Big) \nonumber \\
&+& \sum_{k=1}^{K} W_{k1}(\bbeta^{(t)})\log\Big( w_{k1}(\bbeta^{(t)}) \Big) + \sum_{k=1}^{K} W_{k2}(\bbeta^{(t)})\log\Big( w_{k2}(\bbeta^{(t)}) \Big) \nonumber \\
&+& \sum_{k=1}^{K} W_{k}(\bbeta^{(t)})\log\Big( W_{k}(\bbeta^{(t)}) \Big). \nonumber
\end{eqnarray}

\medskip

\noindent
\textbf{Proof of Theorem 2.} Our aim is to minimize
\begin{equation}
\hat{Q}_{\tau}^{(1)}(\bbeta) 
= \frac{1}{12} - \frac{\tau}{K}\sqrt{\frac{2}{\pi}}\sum_{k=1}^{K}V_{k,\tau} \phi\Big( V_{k,\tau}(Y_{k} - \mathbf{x}_{k}^{T}\bbeta) \Big) +  \frac{1}{K}\sum_{k=1}^{K} \Bigg(\Phi\Big( V_{k,\tau}(Y_{k} - \mathbf{x}_{k}^{T}\bbeta) \Big) - \frac{1}{2} \Bigg)^{2}. \nonumber 
\end{equation}
Minimizing $\hat{Q}_{\tau}^{(1)}(\bbeta)$ is equivalent to maximizing $\tilde{Q}_{\tau}(\bbeta)$, where $\tilde{Q}_{\tau}(\bbeta)$ is defined as
\begin{eqnarray}
\tilde{Q}_{\tau}(\bbeta) 
&=& \sum_{k=1}^{K} V_{k,\tau} \phi\Big( V_{k,\tau}(Y_{k} - \mathbf{x}_{k}^{T}\bbeta) \Big) + \frac{\sqrt{\pi}K}{\tau\sqrt{2}} -  \frac{\sqrt{\pi}}{\tau\sqrt{2}}\sum_{k=1}^{K} \Bigg(\Phi\Big( V_{k,\tau}(Y_{k} - \mathbf{x}_{k}^{T}\bbeta) \Big) - \frac{1}{2} \Bigg)^{2} \nonumber 
\end{eqnarray}
and maximizing $\tilde{Q}_{\tau}(\bbeta)$ is equivalent to minimizing
\begin{eqnarray}
\ell_{\tau}(\bbeta) &=& -\log \tilde{Q}_{\tau}(\bbeta) \nonumber \\
&=& -\log\Bigg[ \sum_{k=1}^{K} \Bigg\{ V_{k,\tau} \phi\Big( V_{k,\tau}(Y_{k} - \mathbf{x}_{k}^{T}\bbeta) \Big) 
 + \frac{\sqrt{\pi}}{\tau\sqrt{2}} -  \frac{\sqrt{\pi}}{\tau\sqrt{2}}\Bigg(\Phi\Big( V_{k,\tau}(Y_{k} - \mathbf{x}_{k}^{T}\bbeta) \Big) - \frac{1}{2} \Bigg)^{2} \Bigg\} \Bigg]. \nonumber 
\end{eqnarray}
Note that we can express $\ell_{\tau}(\bbeta)$ as
$\ell_{\tau}(\bbeta) = -\log( \sum_{k} h_{k}(\bbeta) )$, where 
\begin{eqnarray}
h_{k}(\bbeta) &=& f_{k}(\bbeta) + g_{k}(\bbeta) \nonumber \\
f_{k}(\bbeta) &=& V_{k,\tau} \phi\Big( V_{k,\tau}(Y_{k} - \mathbf{x}_{k}^{T}\bbeta) \Big) \nonumber \\
g_{k}(\bbeta) &=& \frac{\sqrt{\pi}}{\tau\sqrt{2}} - \frac{\sqrt{\pi}}{\tau \sqrt{2}} \Bigg(\Phi\Big( V_{k,\tau}(Y_{k} - \mathbf{x}_{k}^{T}\bbeta) \Big) - \frac{1}{2} \Bigg)^{2} = \frac{\sqrt{\pi}}{\tau\sqrt{2}}\tilde{g}_{k}(\bbeta). \nonumber
\end{eqnarray}
It follows from Lemma \ref{lem:mm2} that
\begin{eqnarray}
\ell_{\tau}(\bbeta)
&\leq&  -\sum_{k=1}^{K} W_{k1}(\bbeta^{(t)})\log\Big( f_{k}(\bbeta) \Big) - \sum_{k=1}^{K} W_{k2}(\bbeta^{(t)})\log\Big( g_{k}(\bbeta) \Big) \nonumber \\
&+& \sum_{k=1}^{K} W_{k1}(\bbeta^{(t)})\log\Big( w_{k1}(\bbeta^{(t)}) \Big) + \sum_{k=1}^{K} W_{k2}(\bbeta^{(t)})\log\Big( w_{k2}(\bbeta^{(t)}) \Big) \nonumber \\
&+& \sum_{k=1}^{K} W_{k}(\bbeta^{(t)})\log\Big( W_{k}(\bbeta^{(t)}) \Big), \nonumber
\end{eqnarray}
and we can simplify the above to
\begin{eqnarray}
\ell_{\tau}(\bbeta)
&\leq&  -\bbeta^{T}\mathbf{X}^{T}\mathbf{W}_{1}\mathbf{V}\mathbf{Y} + \frac{1}{2}\bbeta^{T}\mathbf{X}^{T}\mathbf{W}_{1}\mathbf{V}\mathbf{X}\bbeta - \sum_{k=1}^{K} W_{k2}(\bbeta^{(t)})\log\Big( \tilde{g}_{k}(\bbeta) \Big) \nonumber \\ 
&+& \frac{1}{2}\mathbf{Y}^{T}\mathbf{W}_{1}\mathbf{V}\mathbf{Y} -\sum_{k=1}^{K} W_{k1}(\bbeta^{(t)})\log\Big( V_{k,\tau}/\sqrt{2\pi} \Big) - \sum_{k=1}^{K} W_{k2}(\bbeta^{(t)})\log\Big( \sqrt{\pi}/\tau\sqrt{2} \Big) \nonumber \\
&+& \sum_{k=1}^{K} W_{k1}(\bbeta^{(t)})\log\Big( w_{k1}(\bbeta^{(t)}) \Big) + \sum_{k=1}^{K} W_{k2}(\bbeta^{(t)})\log\Big( w_{k2}(\bbeta^{(t)}) \Big) \nonumber \\
&+& \sum_{k=1}^{K} W_{k}(\bbeta^{(t)})\log\Big( W_{k}(\bbeta^{(t)}) \Big).
\label{eq:mm_first_simplify}
\end{eqnarray}
It then follows from (\ref{eq:mm_first_simplify}) and Lemma \ref{lem:mm1} that
\begin{eqnarray}
\ell_{\tau}(\bbeta)
&\leq&  -\bbeta^{T}\mathbf{X}^{T}\mathbf{W}_{1}\mathbf{V}\mathbf{Y} + \frac{1}{2}\bbeta^{T}\mathbf{X}^{T}\mathbf{W}_{1}\mathbf{V}\mathbf{X}\bbeta + \tilde{G}(\bbeta|\bbeta^{(t)}, \mathbf{w}_{2}) \nonumber \\ 
&+& \frac{1}{2}\mathbf{Y}^{T}\mathbf{W}_{1}\mathbf{V}\mathbf{Y} -\sum_{k=1}^{K} W_{k1}(\bbeta^{(t)})\log\Big( \frac{ V_{k,\tau}}{\sqrt{2\pi} w_{k1}(\bbeta^{(t)})} \Big) - \sum_{k=1}^{K} W_{k2}(\bbeta^{(t)})\log\Big( \frac{ \sqrt{\pi}}{\sqrt{2}\tau w_{k2}(\bbeta^{(t)}) } \Big) \nonumber \\
&+& \sum_{k=1}^{K} W_{k}(\bbeta^{(t)})\log\Big( W_{k}(\bbeta^{(t)}) \Big),
\label{eq:mm_second_simplify}
\end{eqnarray}
where $\tilde{G}(\bbeta|\bbeta^{(t)}, \mathbf{w}_{2})$ is as defined in Lemma \ref{lem:mm2}.
\begin{equation}
\tilde{G}(\bbeta|\bbeta^{(t)}, \mathbf{w}_{2}) = C_{\tilde{G}}^{(t)}(\mathbf{w}_{2}) - \bbeta^{T}\Big[\mathbf{X}^{T}\mathbf{V}^{T}\mathbf{W}_{2}\mathbf{d}(\bbeta^{(t)})
+ \frac{1}{6}\mathbf{X}^{T}\mathbf{V}^{T}\mathbf{W}_{2}\mathbf{V}\mathbf{X}\bbeta^{(t)} \Big]
+ \frac{1}{12}\bbeta^{T}\mathbf{X}^{T}\mathbf{V}^{T}\mathbf{W}_{2}\mathbf{V}\mathbf{X}\bbeta. \nonumber 
\end{equation}

\noindent
Hence, $\ell_{\tau}(\bbeta) \leq H(\bbeta|\bbeta^{(t)})$, where $H(\bbeta|\bbeta^{(t)})$ is defined as
\begin{eqnarray}
H(\bbeta|\bbeta^{(t)})
&=&  -\bbeta^{T}\mathbf{X}^{T}\mathbf{W}_{1}\mathbf{V}\mathbf{Y} + \frac{1}{2}\bbeta^{T}\mathbf{X}^{T}\mathbf{W}_{1}\mathbf{V}\mathbf{X}\bbeta + \tilde{G}(\bbeta|\bbeta^{(t)}, \mathbf{w}_{2}) \nonumber \\ 
&+& \frac{1}{2}\mathbf{Y}^{T}\mathbf{W}_{1}\mathbf{V}\mathbf{Y} -\sum_{k=1}^{K} W_{k1}(\bbeta^{(t)})\log\Big( \frac{ V_{k,\tau}}{\sqrt{2\pi} w_{k1}(\bbeta^{(t)})} \Big) - \sum_{k=1}^{K} W_{k2}(\bbeta^{(t)})\log\Big( \frac{ \sqrt{\pi}}{\sqrt{2}\tau w_{k2}(\bbeta^{(t)}) } \Big) \nonumber \\
&+& \sum_{k=1}^{K} W_{k}(\bbeta^{(t)})\log\Big( W_{k}(\bbeta^{(t)}) \Big).
\label{eq:Hfirst}
\end{eqnarray}
Now, if we look at $H(\bbeta^{(t)}|\bbeta^{(t)})$, we have:
\begin{eqnarray}
H(\bbeta^{(t)}|\bbeta^{(t)})
&=& -\sum_{k=1}^{K} W_{k2}(\bbeta^{(t)})\log(\sqrt{\pi}/\tau\sqrt{2}) -\sum_{k=1}^{K} W_{k1}(\bbeta^{(t)})\log\Big( \frac{f_{k}(\bbeta^{(t)})}{w_{k1}(\bbeta^{(t)})} \Big) - G(\bbeta^{(t)}|\bbeta^{(t)}) \nonumber \\
&+&  \sum_{k=1}^{K} W_{k2}(\bbeta^{(t)})\log\Big( w_{k2}(\bbeta^{(t)}) \Big) + \sum_{k=1}^{K} W_{k}(\bbeta^{(t)})\log\Big( W_{k}(\bbeta^{(t)}) \Big) \nonumber \\
&=& -\sum_{k=1}^{K} W_{k1}(\bbeta^{(t)})\log\Big( \frac{f_{k}(\bbeta^{(t)})}{w_{k1}(\bbeta^{(t)})} \Big) -\sum_{k=1}^{K} W_{k2}(\bbeta^{(t)})\log\Big(\frac{ g_{k}(\bbeta^{(t)})}{  w_{k2}(\bbeta^{(t)}) } \Big) \nonumber \\
&+&  \sum_{k=1}^{K} W_{k}(\bbeta^{(t)})\log\Big( W_{k}(\bbeta^{(t)}) \Big) \nonumber \\
&=& -\sum_{k=1}^{K} W_{k1}(\bbeta^{(t)})\log\Big( h_{k}(\bbeta^{(t)}) \Big) -\sum_{k=1}^{K} W_{k2}(\bbeta^{(t)})\log\Big( h_{k}(\bbeta^{(t)}) \Big) \nonumber \\
&+&  \sum_{k=1}^{K} W_{k}(\bbeta^{(t)})\log\Big( W_{k}(\bbeta^{(t)}) \Big) \nonumber \\
&=& -\sum_{k=1}^{K} [W_{k1}(\bbeta^{(t)}) + W_{k2}(\bbeta^{(t)}]\log\Big( h_{k}(\bbeta^{(t)}) \Big) 
+  \sum_{k=1}^{K} W_{k}(\bbeta^{(t)})\log\Big( W_{k}(\bbeta^{(t)}) \Big) \nonumber \\
&=& -\sum_{k=1}^{K} W_{k}(\bbeta^{(t)})\log\Big( h_{k}(\bbeta^{(t)}) \Big) 
+  \sum_{k=1}^{K} W_{k}(\bbeta^{(t)})\log\Big( W_{k}(\bbeta^{(t)}) \Big) \nonumber \\
&=& -\sum_{k=1}^{K} W_{k}(\bbeta^{(t)})\log\Big( \frac{ h_{k}(\bbeta^{(t)}) }{ W_{k}(\bbeta^{(t)})  } \Big) \nonumber \\
&=& -\sum_{k=1}^{K} W_{k}(\bbeta^{(t)})\log\Big( \sum_{k=1}^{K} h_{k}(\bbeta^{(t)}) \Big) \nonumber \\
&=& -\log\Big( \sum_{k=1}^{K} h_{k}(\bbeta^{(t)}) \Big) \nonumber \\
&=& \ell_{\tau}(\bbeta^{(t)})
\end{eqnarray}
Thus, we have established that  $\ell_{\tau}(\bbeta) \leq H(\bbeta|\bbeta^{(t)})$ and $\ell_{\tau}(\bbeta^{(t)}) = H(\bbeta^{(t)}|\bbeta^{(t)})$. Hence, an iterative scheme defined by $\bbeta^{(t+1)} = \argmin_{\boldsymbol{\beta}} H(\bbeta|\bbeta^{(t)})$ is a valid MM algorithm.

\medskip

\noindent
Because $\tilde{G}(\bbeta|\bbeta^{(t)}, \mathbf{w}_{2})$ in Lemma \ref{lem:mm2} is defined as 
\begin{equation}
\tilde{G}(\bbeta|\bbeta^{(t)}, \mathbf{w}_{2}) = C_{\tilde{G}}^{(t)}(\mathbf{w}_{2}) - \bbeta^{T}\Big[\mathbf{X}^{T}\mathbf{V}^{T}\mathbf{W}_{2}\mathbf{d}(\bbeta^{(t)})
+ \frac{1}{6}\mathbf{X}^{T}\mathbf{V}^{T}\mathbf{W}_{2}\mathbf{V}\mathbf{X}\bbeta^{(t)} \Big]
+ \frac{1}{12}\bbeta^{T}\mathbf{X}^{T}\mathbf{V}^{T}\mathbf{W}_{2}\mathbf{V}\mathbf{X}\bbeta, \nonumber 
\end{equation}
we can simplify $H(\bbeta|\bbeta^{(t)})$ to 
\begin{eqnarray}
H(\bbeta|\bbeta^{(t)})
&=&  - \bbeta^{T}\Big[\mathbf{X}^{T}\mathbf{V}^{T}\mathbf{W}_{1}\mathbf{V}\mathbf{Y} + \mathbf{X}^{T}\mathbf{V}^{T}\mathbf{W}_{2}\mathbf{d}(\bbeta^{(t)})
+ \frac{1}{3}\mathbf{X}^{T}\mathbf{V}^{T}\mathbf{W}_{2}\mathbf{V}\mathbf{X}\bbeta^{(t)} \Big] \nonumber \\
&+& \frac{1}{2}\bbeta^{T}\Big( \mathbf{X}^{T}\mathbf{V}^{T}\mathbf{W}_{1}\mathbf{V}\mathbf{X} + \frac{1}{3}\mathbf{X}^{T}\mathbf{V}^{T}\mathbf{W}_{2}\mathbf{V}\mathbf{X} \Big) \bbeta  \nonumber \\ 
&+& C_{\tilde{G}}^{(t)}(\mathbf{w}_{2}) + \frac{1}{2}\mathbf{Y}^{T}\mathbf{V}^{T}\mathbf{W}_{1}\mathbf{V}\mathbf{Y} -\sum_{k=1}^{K} W_{k1}(\bbeta^{(t)})\log\Big( \frac{ V_{k,\tau}}{\sqrt{2\pi} w_{k1}(\bbeta^{(t)})} \Big) \nonumber \\
&-& \sum_{k=1}^{K} W_{k2}(\bbeta^{(t)})\log\Big( \frac{ \sqrt{\pi}}{\sqrt{2}\tau w_{k2}(\bbeta^{(t)}) } \Big) 
+ \sum_{k=1}^{K} W_{k}(\bbeta^{(t)})\log\Big( W_{k}(\bbeta^{(t)}) \Big).
\label{eq:Hsimplify}
\end{eqnarray}
Minimizing the above with respect to $\bbeta$ yields the following MM algorithm update:
\begin{equation}
\bbeta^{(t+1)} = \Bigg(\mathbf{X}^{T}\mathbf{V}^{T}(\mathbf{W}_{1}^{(t)} + \tfrac{1}{3}\mathbf{W}_{2}^{(t)})\mathbf{V}\mathbf{X}\Bigg)^{-1}\Big( \mathbf{X}^{T}\mathbf{V}^{T}\{\mathbf{W}_{1}^{(t)}\mathbf{V}\mathbf{Y}  
+ \mathbf{W}_{2}^{(t)}\mathbf{d}(\bbeta^{(t)})
+ \tfrac{1}{3}\mathbf{W}_{2}^{(t)}\mathbf{V}\mathbf{X}\bbeta^{(t)} \} \Big).
\nonumber 
\end{equation}

\section{Data Cleaning Steps, Additional Figures and Tables}

\subsection{Data Cleaning Steps}

For our analysis in Section 5 of the main manuscript, we used kindergarten test results available at \url{https://nces.ed.gov/ecls/dataproducts.asp}.

\medskip

\noindent
After the initial pre-processing, the data should have $109,044$ observations made on $4207$ schools.
From the raw data, we created the summary data used in our analysis through the following steps:
\begin{enumerate}
\item
We first retained only schools that had at least one observation made in Spring 2011. (after this, you should have 18174 observations from 1308 schools)
\item
Among all observations from these 1308 schools, we removed observations where either: the School Locale or School Type (public or private) was missing, OR the total number of observations from the school was less than 5. (after this, you should have 17106 observations from 911 schools)
\item
Among these 911 schools, we only kept schools that had 5 or more observed student test scores and 1 or more observed
teacher education levels. (after this, you should have 862 schools).
\item
For each variable used in the analysis, compute the sample mean within each school. The standard errors were
found by computing the sample standard deviation of math scores within each school and dividing by the square root of the number of observations in that school.
\end{enumerate}

\begin{figure}[h!]
\centering
\includegraphics[width=\textwidth]{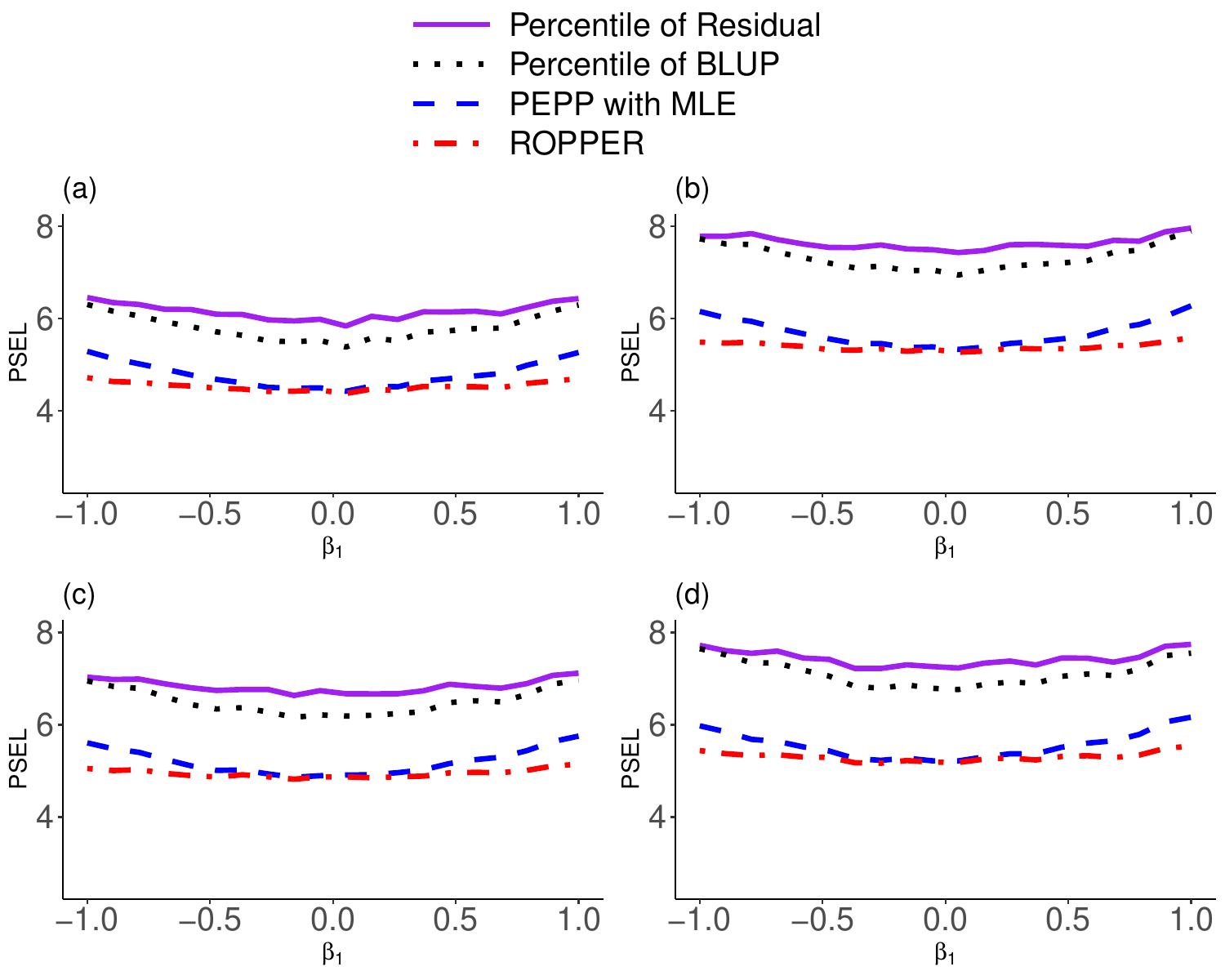}
\caption{Percentile squared-error loss (PSEL) for different ranking methods. Outcomes $Y_{k}$ were simulated using the unmodeled latent subgroup simulation setting described in Section 4.1. Panels: (a) $\varv_k \sim N(0,\tau^2)$ (b) $\varv_k=\sqrt{3}\tau T_k$ where $T_k \sim t(df=3)$ (c) $\varv_k \sim \textrm{Exp}(\mu=1)$ (d) $\varv_k \sim \textrm{Gamma}(\alpha=2,\beta=\sqrt{2})$. $\beta_1$: effect size of unmeasured covariate. PEPP: Posterior Expected Population Percentile. ROPPER: Ranking-Optimized Population Posterior Expected peRcentiles. PSEL values are based on 1000 simulation iterations.}
\label{fig:other_distribution_simresults}
\end{figure}

\begin{figure}[h!]
\centering
\includegraphics[width=\textwidth]{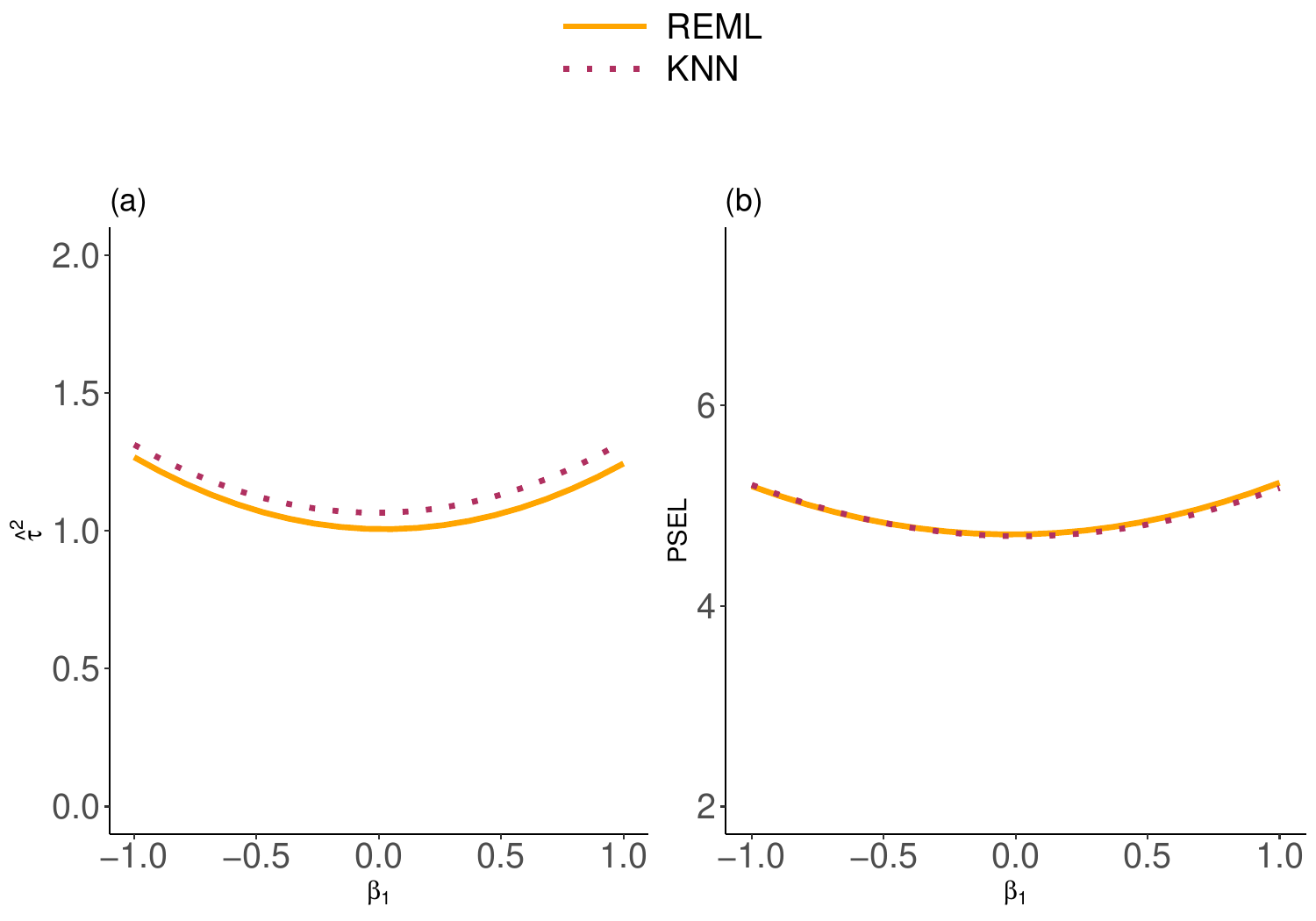}
\caption{Simulation study based on emulating the real-world education data. Panels: (a) Average $\hat{\tau}^2$ value and (b) Percentile squared-error loss (PSEL) for the proposed ROPPER (Ranking-Optimized Population Posterior Expected peRcentiles) method using the Restricted Maximum Likelihood (REML) versus K-nearest neighbor (KNN) $\tau^2$ estimation procedures. The true $\tau^2$ value is 1.0. $\beta_1$: effect size of unmeasured covariate. Based on 1000 simulation iterations.}
\label{fig:knn_compare}
\end{figure}

\begin{figure}[h!]
\centering
\includegraphics[width=\textwidth]{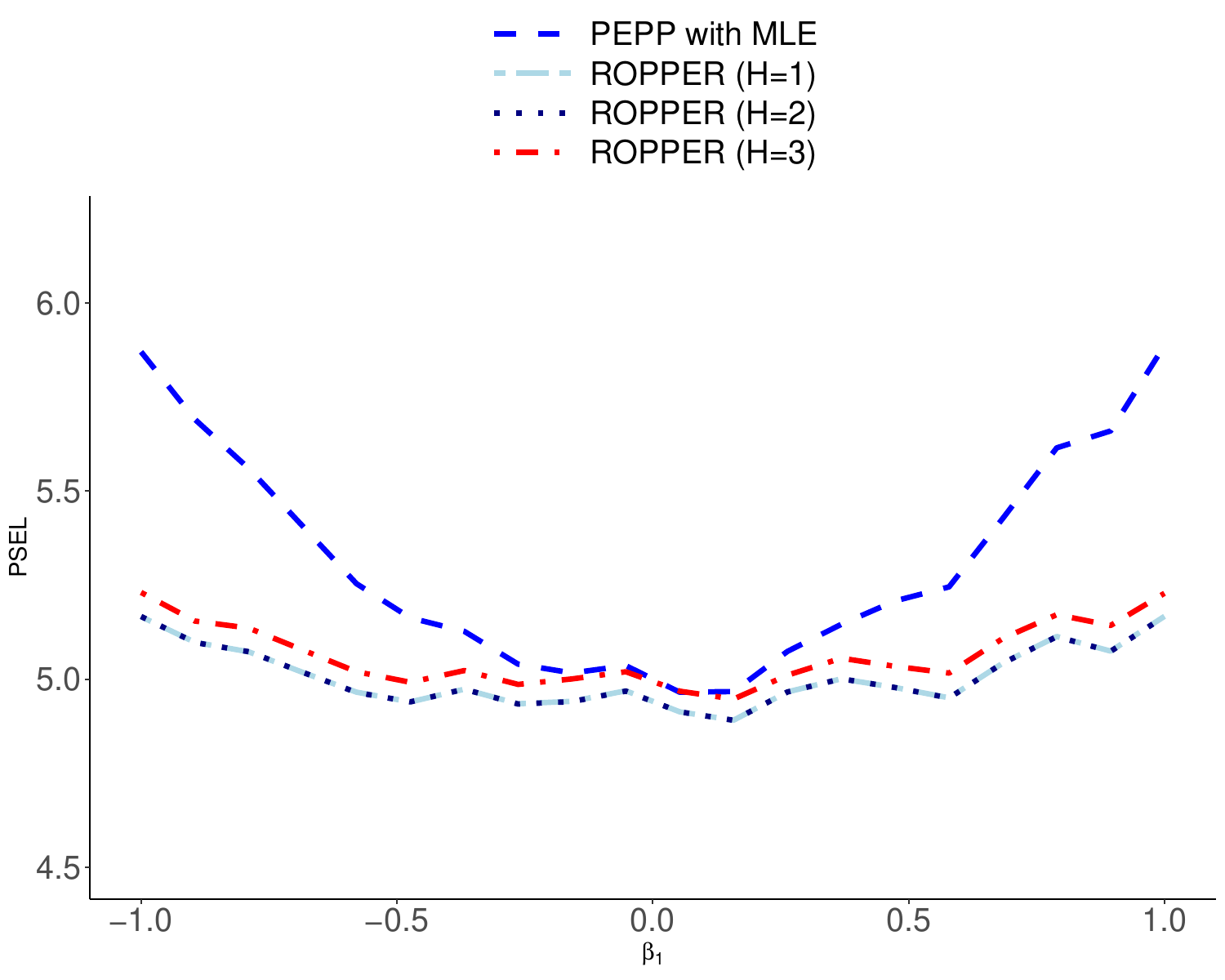}
\caption{Percentile squared-error loss (PSEL) for ROPPER using different choices of the approximation order $H$. Approximation orders $H=1$, $H=2$, and $H=3$ are considered. Outcomes $Y_{k}$ were simulated using the unmodeled latent subgroup simulation setting described in Section 4.1. Simulation design parameters were set to the following values: $K = 50$, $\sigma^{2} = 5$, and $\tau^{2} = 0.75$. PEPP: Posterior Expected Population Percentile. ROPPER: Ranking-Optimized Population Posterior Expected peRcentiles. PSEL values are based on 1000 simulation iterations.}
\label{fig:differentH}
\end{figure}

\end{document}